\documentclass[fleqn,usenatbib,useAMS]{mnras}
 \usepackage{graphicx}
 \usepackage{amsmath}
 \usepackage{amssymb}
 \usepackage[T1]{fontenc}
 \usepackage{ae,aecompl}
 \usepackage{txfonts}
 \usepackage{enumerate}

 \title[The recurrent dynamics of 469219]
       {Asteroid (469219) 2016~HO$_{3}$, the smallest and closest Earth quasi-satellite}

 \author[C. de la Fuente Marcos and R. de la Fuente Marcos]
        {C.~de~la~Fuente~Marcos\thanks{E-mail: carlosdlfmarcos@gmail.com}
         and
         R. de la Fuente Marcos \\
         Apartado de Correos 3413, E-28080 Madrid, Spain}
 \date{Accepted 2016 August 4.
       Received 2016 August 3;
       in original form 2016 June 12}
 \pubyear{2016}
 
\begin{document}
  \label{firstpage}
  \pagerange{\pageref{firstpage}--\pageref{lastpage}}
  \maketitle

  \begin{abstract}
     A number of Earth co-orbital asteroids experience repeated transitions 
     between the quasi-satellite and horseshoe dynamical states. Asteroids 
     2001~GO$_{2}$, 2002~AA$_{29}$, 2003~YN$_{107}$ and 2015~SO$_{2}$ are 
     well-documented cases of such a dynamical behaviour. These transitions 
     depend on the gravitational influence of other planets, owing to the 
     overlapping of a multiplicity of secular resonances. Here, we show 
     that the recently discovered asteroid (469219) 2016~HO$_{3}$ is a 
     quasi-satellite of our planet ---the fifth one, joining the ranks of 
     (164207)~2004~GU$_{9}$, (277810)~2006~FV$_{35}$, 2013~LX$_{28}$ and 
     2014~OL$_{339}$. This new Earth co-orbital also switches repeatedly 
     between the quasi-satellite and horseshoe configurations. Its current 
     quasi-satellite episode started nearly 100 yr ago and it will end in 
     about 300 yr from now. The orbital solution currently available for 
     this object is very robust and our full $N$-body calculations show 
     that it may be a long-term companion (time-scale of Myr) to our planet. 
     Among the known Earth quasi-satellites, it is the closest to our 
     planet and as such, a potentially accessible target for future in situ 
     study. Due to its presumably lengthy dynamical relationship with the 
     Earth and given the fact that at present and for many decades this 
     transient object remains well positioned with respect to our planet, 
     the results of spectroscopic studies of this small body, 26--115~m, 
     may be particularly useful to improve our understanding of the origins 
     ---local or captured--- of Earth's co-orbital asteroid population. The
     non-negligible effect of the uncertainty in the value of the mass of 
     Jupiter on the stability of this type of co-orbitals is also briefly 
     explored.
  \end{abstract}


  \section{Introduction}
     The study of the orbital dynamics of small-body populations in the Solar system has multiple practical applications ranging from the
     computation of impact and cratering rates on to planetary bodies (see e.g. Michel \& Morbidelli 2007) to selection of spacecraft 
     accessible targets (see e.g. Adamo et al. 2010). In the particular case of the near-Earth objects (NEOs), the analysis of their 
     present-day orbital architecture, their past orbital evolution, and their future scattering events has made possible to improve 
     steadily our understanding of the dynamical properties of the immediate neighbourhood of our planet. 

     In general and for a given planetary host, the small-body populations subjected to direct gravitational interaction with the host are 
     either primordial or transient. Transient small-body populations consistently dominate, but primordial dynamical groups are known to 
     exist. Primordial dynamical groups in direct gravitational interaction with a planetary host are captured permanently in resonance and
     they may have formed near their present location. Within this context, perhaps the most relevant resonance is the 1:1 mean motion 
     resonance that makes objects go around a central star in almost exactly one planetary orbital period. These interesting bodies are 
     termed co-orbitals even if their orbits only resemble that of the host planet in terms of orbital period but they are very eccentric 
     and/or inclined (Morais \& Morbidelli 2002). Frozen in orbital parameter space, primordial co-orbital small bodies are the remnants of 
     the physical and dynamical processes that molded their planetary host and this fact makes them particularly interesting when attempting 
     to reconstruct the conditions reigning at that moment. Well-known examples of these populations are the Trojans of Jupiter (see e.g. Di 
     Sisto, Ramos \& Beaug\'e 2014; Wong, Brown \& Emery 2014), Mars (see e.g. Marzari et al. 2002; Connors et al. 2005; Scholl, Marzari \& 
     Tricarico 2005; \'Cuk, Christou \& Hamilton 2015) or Neptune (see e.g. Parker 2015; Gerdes et al. 2016). In addition to primordial 
     co-orbitals, objects from the planet-crossing small-body populations can undergo temporary co-orbital capture giving rise to transient 
     co-orbital populations. In a multi-planet environment, the orbital evolution of transient co-orbitals (including capture and ejection) 
     is mostly controlled by secular resonances induced by planets other than the host. 

     Small bodies are classified dynamically (for instance as Atens, Apollos, Centaurs, etc) based on their present-day orbital solutions. 
     The case of co-orbitals, both primordial and transient, is however more complicated. Having a particular set of orbital elements at 
     present time is not enough to claim a co-orbital relationship with a host; a representative set of statistically compatible orbits must 
     be integrated forward and backwards in time to show that the dynamical evolution of the object over a reasonable amount of time is also 
     consistent with being locked in a 1:1 mean motion resonance with the host. Therefore, co-orbital objects are identified indirectly after 
     investigating numerically their orbital evolution. 

     Minor bodies are classified as co-orbitals of a given host after studying the behaviour of a critical angle. The variable of interest
     here is the relative mean longitude, $\lambda_{\rm r}$, or difference between the mean longitude of the object and that of its host. In 
     celestial mechanics, the mean longitude of an object ---planet or minor body--- is given by $\lambda=M+\Omega+\omega$, where $M$ is the 
     mean anomaly, $\Omega$ is the longitude of the ascending node, and $\omega$ is the argument of perihelion (see e.g. Murray \& Dermott 
     1999). The relative mean longitude of a non-resonant object with respect to a second body circulates; i.e. $\lambda_{\rm r}\in(0, 
     360)\degr$ with all the values being equally probable. However, if an object is locked in a 1:1 mean motion resonance with a host, the 
     critical or resonant angle $\lambda_{\rm r}$ librates or oscillates around a certain, well-defined value. Although the critical value 
     is a function of the orbital eccentricity and inclination of the objects involved (Namouni, Christou \& Murray 1999; Namouni \& Murray 
     2000), 0\degr, $\pm$60\degr or 180\degr are often cited in the literature as the signposts of co-orbital behaviour (see e.g. Murray \& 
     Dermott 1999).

     There are three elementary or primary co-orbital configurations: quasi-satellite or retrograde satellite, Trojan or tadpole, and 
     horseshoe. However, in a multi-planet environment, the co-orbital dynamics experienced by an object may show a surprisingly high level 
     of complexity as compound states (see e.g. Morais \& Morbidelli 2002) and recurrent transitions between the three elementary co-orbital
     dynamical states (Namouni et al. 1999; Namouni \& Murray 2000) are feasible. Out of the three primary co-orbital configurations, the 
     quasi-satellite dynamical state is the rarest and it is characterized by the libration of $\lambda_{\rm r}$ about 0$^{\circ}$. This 
     value of the resonant angle is independent from the nature of the orbital eccentricity and inclination of the objects involved. 
     Although a bona fide quasi-satellite appears to travel around its host when its motion is viewed in a frame of reference rotating with 
     the host (see e.g. Fig. \ref{qs}), the object is not gravitationally bound to it and describes complex, drifting loops as observed from 
     the host (see e.g. Fig. \ref{qsradec}). 

     The term `quasi-satellite' was brought before and popularized among the astronomical community by Mikkola \& Innanen (1997). However, 
     the associated concept had been first studied by Jackson (1913) and its energy balance was initially explored by H\'enon (1969), who 
     coined the term `retrograde satellites' to refer to them, although this term is seldom used nowadays. The details of this interesting
     co-orbital configuration were further studied by Szebehely (1967), Broucke (1968), Benest (1976, 1977), Dermott \& Murray (1981), Kogan 
     (1989) and Lidov \& Vashkov'yak (1993, 1994a,b); the stability of the quasi-satellite dynamical state has been studied by Mikkola et 
     al. (2006), Sidorenko et al. (2014) and Pousse, Robutel \& Vienne (2016). Most of the results presented in the works cited have been 
     obtained within the simplified framework of the restricted elliptic three-body problem. However, the interest in this orbital 
     configuration is far from strictly theoretical. Quasi-satellite configurations may have played a role in the origin of the Earth--Moon 
     system (Kortenkamp \& Hartmann 2016); they have also been found numerically (de la Fuente Marcos, de la Fuente Marcos \& Aarseth 2016) 
     within the context of the Planet Nine hypothesis (Batygin \& Brown 2016). 

     Although the existence of quasi-satellites was predicted over a century ago, the first minor body to be confirmed to follow a
     quasi-satellite trajectory, in this case with respect to Venus, was 2002~VE$_{68}$ in 2004 (Mikkola et al. 2004; de la Fuente Marcos \& 
     de la Fuente Marcos 2012a). Asteroid 2002~VE$_{68}$ is so far the only known quasi-satellite of Venus. Among the inner planets, the 
     Earth hosts four confirmed quasi-satellite companions: (164207) 2004~GU$_{9}$ (Connors et al. 2004; Mikkola et al. 2006; Wajer 2010), 
     (277810) 2006~FV$_{35}$ (Wiegert et al. 2008; Wajer 2010), 2013~LX$_{28}$ (Connors 2014), and 2014~OL$_{339}$ (de la Fuente Marcos \& 
     de la Fuente Marcos 2014, 2016c). In the main asteroid belt, objects in this co-orbital configuration have been found pursuing Ceres 
     and Vesta (Christou 2000b; Christou \& Wiegert 2012). Jupiter appears to host the largest known population of quasi-satellites in the
     Solar system with at least six, including asteroids and comets (Kinoshita \& Nakai 2007; Wajer \& Kr\'olikowska 2012). Saturn (Gallardo 
     2006) and Neptune (de la Fuente Marcos \& de la Fuente Marcos 2012c) have one quasi-satellite each. Even dwarf planet Pluto has at 
     least one (de la Fuente Marcos \& de la Fuente Marcos 2012b, but see Porter et al. 2016). 

     Therefore, our planet may come in second place regarding number of quasi-satellite companions. Extensive numerical simulations show 
     that the four known Earth quasi-satellites ---164207, 277810, 2013~LX$_{28}$ and 2014~OL$_{339}$--- are of transient nature (see the 
     review in de la Fuente Marcos \& de la Fuente Marcos 2014, 2016c). No candidates to being primordial quasi-satellites have been 
     identified yet, but Kortenkamp (2005) has argued that early in the history of the Solar system, 5 to 20 per cent of planetesimals 
     scattered by any given planet may have become quasi-satellites. At present time, large amounts of interplanetary dust particles are 
     also being temporarily trapped in Earth's quasi-satellite resonance (Kortenkamp 2013).

     Although a quasi-satellite is not a true satellite, it may become one. Co-orbitals that follow orbits similar to that of the host body 
     experience slow encounters with it. In the particular case of planetary hosts and during a slow encounter, the relative velocity 
     between co-orbital and planet can be so slow that the planetocentric energy may become negative. Under these circumstances, the object 
     can experience a temporary satellite capture in strict sense. This theoretical possibility was confirmed dramatically after careful 
     analysis of the orbital evolution of 2006~RH$_{120}$, a transient co-orbital that stayed as natural satellite of our planet for about a 
     year starting in 2006 June (Kwiatkowski et al. 2009; Granvik, Vaubaillon \& Jedicke 2012). Such transient natural satellites are often 
     referred to in the literature as mini-moons (Granvik et al. 2012; Bolin et al. 2014) and they may in some cases collide with the 
     planetary host (Clark et al. 2016) as it happened in July 1994 with comet D/Shoemaker-Levy 9 and Jupiter (Carusi, Marsden \& Valsecchi 
     1994; Benner \& McKinnon 1995; Kary \& Dones 1996). In addition, capture of quasi-satellites could lead to the production of irregular 
     satellites of the Jovian planets (see e.g. Jewitt \& Haghighipour 2007).

     On a more practical side, Earth co-orbitals ---quasi-satellites included--- are also interesting targets for future sample return 
     missions and other outer space activities (see e.g. Lewis 1996; Stacey \& Connors 2009; Elvis 2012, 2014; Garc\'{\i}a Y\'arnoz, Sanchez 
     \& McInnes 2013; Harris \& Drube 2014). A number of these NEOs are relatively easy to access from the Earth because they have both 
     short perigee distances and low orbital inclinations. Consistently, they have been made part of the Near-Earth Object Human Space 
     Flight Accessible Targets Study (NHATS)\footnote{http://neo.jpl.nasa.gov/nhats/} list (Abell et~al. 2012a,b). In general, 
     quasi-satellites do not experience flybys with our planet as close as those observed for other co-orbital types but, in terms of 
     average distance from our planet (see e.g. Fig. \ref{qsradecperi}), some of them ---e.g. 164207--- tend to remain accessible for 
     comparatively lengthy periods of time, making the scheduling of a putative sample return mission easier.

     Although our planet is perhaps only second to Jupiter regarding number of quasi-satellites, horseshoe librators not quasi-satellites 
     dominate the known population of Earth co-orbitals (de la Fuente Marcos \& de la Fuente Marcos 2016a,b). These objects follow horseshoe 
     orbits (see e.g.  Murray \& Dermott 1999) such as the value of $\lambda_{\rm r}$ oscillates about 180\degr with an amplitude wider than 
     180\degr, often enclosing $\pm60\degr$. A number of Earth co-orbitals that at present pursue horseshoe paths ---2001~GO$_{2}$, 
     2002~AA$_{29}$, 2003~YN$_{107}$ and 2015~SO$_{2}$--- experience repeated transitions to and from the quasi-satellite dynamical state 
     (Brasser et al. 2004; de la Fuente Marcos \& de la Fuente Marcos 2016a). This orbital behaviour was first predicted theoretically by 
     Namouni (1999). Here, we show that the recently discovered minor body (469219) 2016~HO$_{3}$ is a quasi-satellite of the Earth that 
     also switches repeatedly between the quasi-satellite and horseshoe configurations. The object was originally selected as a co-orbital 
     candidate because of its small relative semimajor axis, $|a - a_{\rm Earth}| \sim$ 0.0005 au; extensive $N$-body calculations confirm 
     its current Earth quasi-satellite dynamical status. This paper is organized as follows. In Section 2, we briefly outline the details of 
     our numerical model. Section 3 focuses on 469219 and its orbital evolution. Section 4 explores the role of the uncertainty in the mass 
     of Jupiter on the assessment of the stability of current Earth quasi-satellites. Section 5 singles out 469219 as a suitable candidate 
     to perform spectroscopic observations. Our results are discussed in Section 6. Section 7 summarizes our conclusions.    

  \section{Numerical model}
     As pointed out in the previous section, the identification of co-orbital objects is not based on the present-day values of their 
     orbital parameters alone, but on the statistical analysis of the results of large sets of numerical integrations. Aiming at studying 
     the orbital evolution of (469219) 2016~HO$_{3}$ and following the steps outlined in de la Fuente Marcos \& de la Fuente Marcos (2012a, 
     2015b), extensive $N$-body calculations have been performed. These numerical integrations have been carried out using the Hermite 
     scheme (Makino 1991; Aarseth 2003). The standard version of the $N$-body code used in this study is publicly available from S.~J. 
     Aarseth's web site.\footnote{http://www.ast.cam.ac.uk/$\sim$sverre/web/pages/nbody.htm} As explained in detail in de la Fuente Marcos 
     \& de la Fuente Marcos (2012a), results from this code compare well with those from Laskar et al. (2011) among others. At the end of 
     the simulations, the relative errors in the total energy are $< 1 \times 10^{-14}$ and those in the total angular momentum are several 
     orders of magnitude smaller. Our integrations have two main ingredients: initial conditions and physical model. 

     In the case of a nominal orbit, our initial conditions (positions and velocities in the barycentre of the Solar system) have been 
     provided directly by the JPL \textsc{horizons}\footnote{http://ssd.jpl.nasa.gov/?horizons} system (Giorgini et al. 1996; Giorgini \& 
     Yeomans 1999; Giorgini, Chodas \& Yeomans 2001). They are relative to the JD TDB (Julian Date, Barycentric Dynamical Time) epoch 
     2457600.5 (2016-July-31.0), which is the $t$ = 0 instant in the figures. These initial conditions, for both planets and minor bodies, 
     are based on the DE405 planetary orbital ephemerides (Standish 1998). The initial conditions for all the control orbits of 469219 are 
     based on the available orbital solution (see Section 3.5 for details).

     The physical model includes the perturbations from the eight major planets, the Moon, the barycentre of the Pluto-Charon system, and 
     the three largest asteroids. An example of a typical input file with details can be found in the appendix (table 9) of de la Fuente 
     Marcos, de la Fuente Marcos \& Aarseth (2015). Our physical model does not include non-gravitational forces, relativistic or 
     oblateness terms in the integrated equations of motion. We have neglected the effect of the Yarkovsky and 
     Yarkovsky--O'Keefe--Radzievskii--Paddack (YORP) effects (see e.g. Bottke et al. 2006), but ignoring these effects has no relevant 
     impact on the evaluation of the present-day dynamical status of 469219. Not including these non-gravitational forces in the  
     calculations may affect both the reconstruction of the dynamical past of this object and any predictions made regarding its future 
     orbital evolution. However, accurate modelling of the Yarkovsky force requires relatively precise knowledge of the physical properties 
     ---such as rotation rate, albedo, bulk density, surface conductivity or emissivity--- of the objects under study, which is not the case 
     here (see Section 3.1). On the other hand, the effects derived from these forces may be unimportant when objects are tumbling or in 
     chaotic rotation, and NEOs this small often are. Effects resulting from the theory of general relativity are insignificant for objects 
     following orbits like that of 469219 (see e.g. Benitez \& Gallardo 2008). The role of the oblateness of the Earth can be neglected for 
     cases like the one studied here ---see the analysis in Dmitriev, Lupovka \& Gritsevich (2015) for the Chelyabinsk superbolide.  

  \section{Asteroid (469219) 2016~HO$_{3}$, an Apollo quasi-satellite}
     Here, we show the data available for this recently discovered NEO and study both its short- and long-term orbital evolution. Emphasis
     is made on the statistical robustness of our results.

     \subsection{Data}
        Asteroid (469219) 2016~HO$_{3}$ was discovered on 2016 April 27 by B. Gibson, T. Goggia, N. Primak, A. Schultz and M. Willman 
        observing with the 1.8-m Ritchey-Chretien telescope of the Pan-STARRS Project (Kaiser et al. 2004) from Haleakala at an apparent 
        magnitude $w$ of 21.5 (Mastaler et al. 2016).\footnote{http://www.minorplanetcenter.net/mpec/K16/K16H63.html} Its absolute 
        magnitude, $H$ = 24.1 (assumed $G$ = 0.15), suggested a diameter in the range 26--115~m for an assumed albedo in the range 
        0.60--0.03. Additional observations led to an eventual improvement of its original orbital solution (Schwartz et al. 
        2016).\footnote{http://www.minorplanetcenter.net/mpec/K16/K16K07.html}$^{,}$\footnote{http://www.minorplanetcenter.net/mpec/K16/K16L96.html} 
        P. Vere\v{s} of the Pan-STARRS Project Team found precovery images acquired in 2011, 2012, 2013, 2014 and 2015. S. Deen found 
        additional precovery images from the Apache Point-Sloan Digital Sky Survey (Loveday et al. 1998) acquired in 
        2004.\footnote{http://www.minorplanetcenter.net/mpec/K16/K16L07.html} It is not often that observations from multiple oppositions of 
        a small NEO are recovered shortly after the actual discovery. The currently available orbital solution for this object (see Table 
        \ref{elements}) is based on 80 observations spanning a data-arc of 4468 d or 12.23 yr, from 2004 March 17 to 2016 June 10, its 
        residual rms amounts to 0.28 arcsec. This orbital determination places 469219 among the group of NEOs with robust orbital 
        solutions. With a value of the semimajor axis $a$ = 1.0012 au, this NEO is an Apollo asteroid moving in a low-eccentricity, $e$ = 
        0.10, low-inclination, $i$ = 7\fdg77, orbit that keeps the motion of this object confined to the neighbourhood of the Earth--Moon 
        system, without experiencing any close approaches to other planets. Its Minimum Orbit Intersection Distance (MOID) with our planet 
        is 0.0345~au. As a recent discovery, little else besides its orbit and presumed size is known about this minor 
        body.\footnote{http://www.jpl.nasa.gov/news/news.php?feature=6537} 
%
%
         \begin{table}
          \fontsize{8}{11pt}\selectfont
          \tabcolsep 0.10truecm
          \caption{Heliocentric Keplerian orbital elements of (469219) 2016~HO$_{3}$ used in this study. The orbital solution is based on 80 
                   observations spanning a data-arc of 4468 d or 12.23 yr, from 2004 March 17 to 2016 June 10. Values include the 
                   1$\sigma$ uncertainty. The orbit has been computed at epoch JD 2457600.5 that corresponds to 00:00:00.000 TDB on 2016 
                   July 31 (J2000.0 ecliptic and equinox. Source: JPL Small-Body Database.)
                  }
          \begin{tabular}{lcc}
           \hline
            Semimajor axis, $a$ (au)                          & = & 1.001229935$\pm$0.000000003 \\
            Eccentricity, $e$                                 & = & 0.1041429$\pm$0.0000005 \\
            Inclination, $i$ (\degr)                          & = & 7.77140$\pm$0.00004 \\
            Longitude of the ascending node, $\Omega$ (\degr) & = & 66.51326$\pm$0.00004 \\
            Argument of perihelion, $\omega$ (\degr)          & = & 307.22765$\pm$0.00007 \\
            Mean anomaly, $M$ (\degr)                         & = & 297.53211$\pm$0.00010 \\
            Perihelion, $q$ (au)                              & = & 0.8969589$\pm$0.0000005 \\
            Aphelion, $Q$ (au)                                & = & 1.105500933$\pm$0.000000003 \\
            Absolute magnitude, $H$ (mag)                     & = & 24.1$\pm$0.5 \\
           \hline
          \end{tabular}
          \label{elements}
         \end{table}
%
%

%
%
     \begin{figure}
       \centering
        \includegraphics[width=\linewidth]{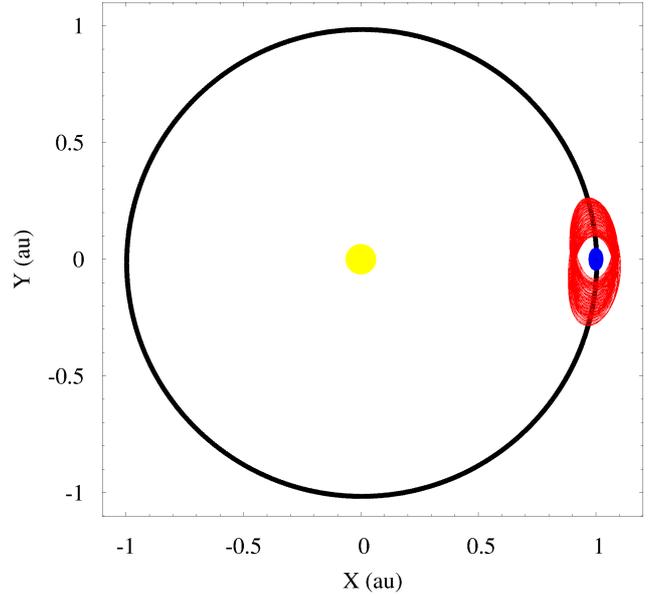}
        \caption{The motion of (469219) 2016~HO$_{3}$ over the time interval ($-$100, 200) yr according to the nominal orbit in Table 
                 \ref{elements} is displayed projected on to the ecliptic plane in a heliocentric frame of reference that rotates with the 
                 Earth. The orbit and position of our planet are also indicated. All the investigated control orbits exhibit the same 
                 behaviour within this time interval.
                }
        \label{qs}
     \end{figure}
%
%
     \subsection{Quasi-satellite of the Earth}
        Fig. \ref{qs} shows the motion of (469219) 2016~HO$_{3}$ (in red, nominal orbit in Table \ref{elements}) over the time interval 
        ($-$100, 200) yr projected on to the ecliptic plane as seen in a frame of reference centred at the Sun and rotating with the Earth. 
        All the investigated control orbits of this small body exhibit the same resonant behaviour within this time interval; the observed 
        evolution is consistent with the one described in Mikkola et al. (2006), Sidorenko et al. (2014) or Pousse et al. (2016). This 
        result shows that 469219 is a co-orbital that currently follows a quasi-satellite orbit around our planet. It joins the ranks of 
        (164207) 2004~GU$_{9}$, (277810)~2006~FV$_{35}$, 2013~LX$_{28}$, and 2014~OL$_{339}$ as the fifth quasi-satellite of the Earth. It 
        is also the smallest one by 1.5~mag in $H$; the second smallest Earth quasi-satellite is 2014~OL$_{339}$ with $H$ = 22.6~mag. 

\begin{figure}
       \centering
        \includegraphics[width=\linewidth]{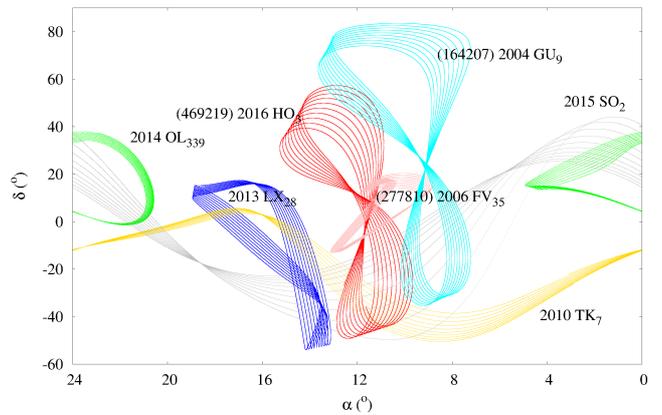}
        \caption{Apparent motion in geocentric equatorial coordinates of the known quasi-satellites, 2010~TK$_{7}$, a Trojan, and 
                 2015~SO$_{2}$, a horseshoe librator, over the time range (0, 10) yr. Earth's quasi-satellites describe complex, drifting 
                 loops as observed from our planet. Both (164207) 2004~GU$_{9}$ and (469219) 2016~HO$_{3}$ trace figure-eight paths. 
                }
        \label{qsradec}
     \end{figure}

\begin{figure}
       \centering
        \includegraphics[width=\linewidth]{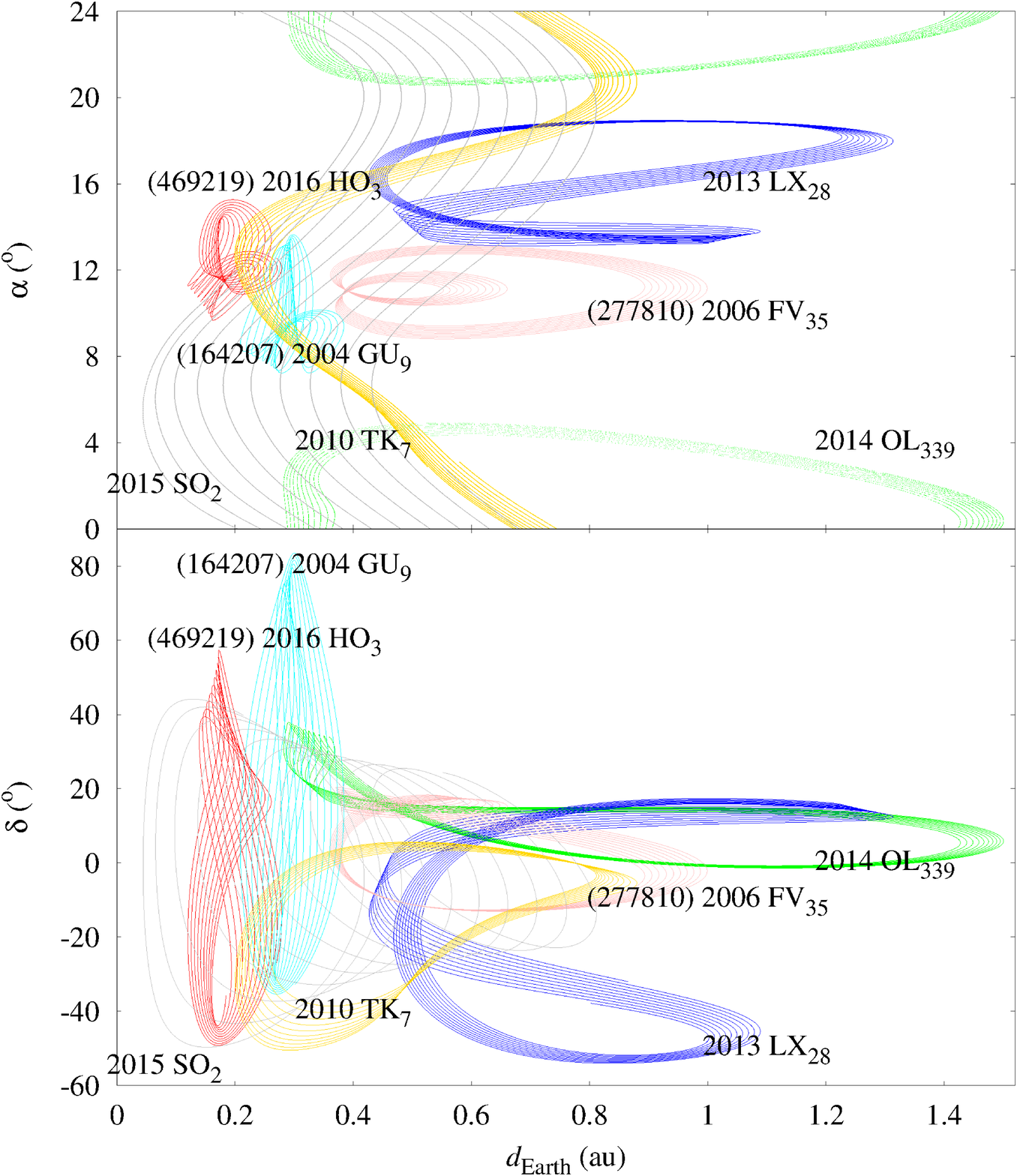}
        \caption{Apparent motion in geocentric equatorial coordinates of the objects in Fig. \ref{qsradec} as a function of their distance 
                 from the Earth over the time range (0, 10) yr. Colours as in Fig. \ref{qsradec}. 
                }
        \label{qsradecperi}
     \end{figure}

        Fig. \ref{qsradec} shows that 469219 describes a shifting figure-eight retrograde path ---as 164207 does--- when viewed from our 
        planet over the course of a sidereal year. When compared with other Earth co-orbitals (2010~TK$_{7}$ and 2015~SO$_{2}$), 
        quasi-satellites delineate very conspicuous paths in the sky, very different from those of gravitationally bound satellites like the 
        Moon or other unbound co-orbitals like 2010~TK$_{7}$ and 2015~SO$_{2}$. When the motion of the known quasi-satellites is plotted as 
        a function of their distance from the Earth (see Fig. \ref{qsradecperi}), it becomes clear why 469219 has a better orbital solution 
        than several other co-orbitals in spite of its small size: its average distance from the Earth is $\sim$0.2 au and never goes beyond 
        0.3~au during its current quasi-satellite episode. Only 164207 exhibits a comparable behaviour in terms of distance from the Earth. 
        Most known quasi-satellites have a wide difference between their perigee and apogee distances; asteroid 2014~OL$_{339}$ is the most 
        extreme example (see Fig. \ref{qsradecperi}). The behaviour observed in Fig. \ref{qsradecperi} makes both 164207 and 469219 
        attractive targets regarding affordable accessibility from our planet in terms of scheduling.
%
%
%
%
%
%
%
%

%
%
     \begin{figure}
       \centering
        \includegraphics[width=\linewidth]{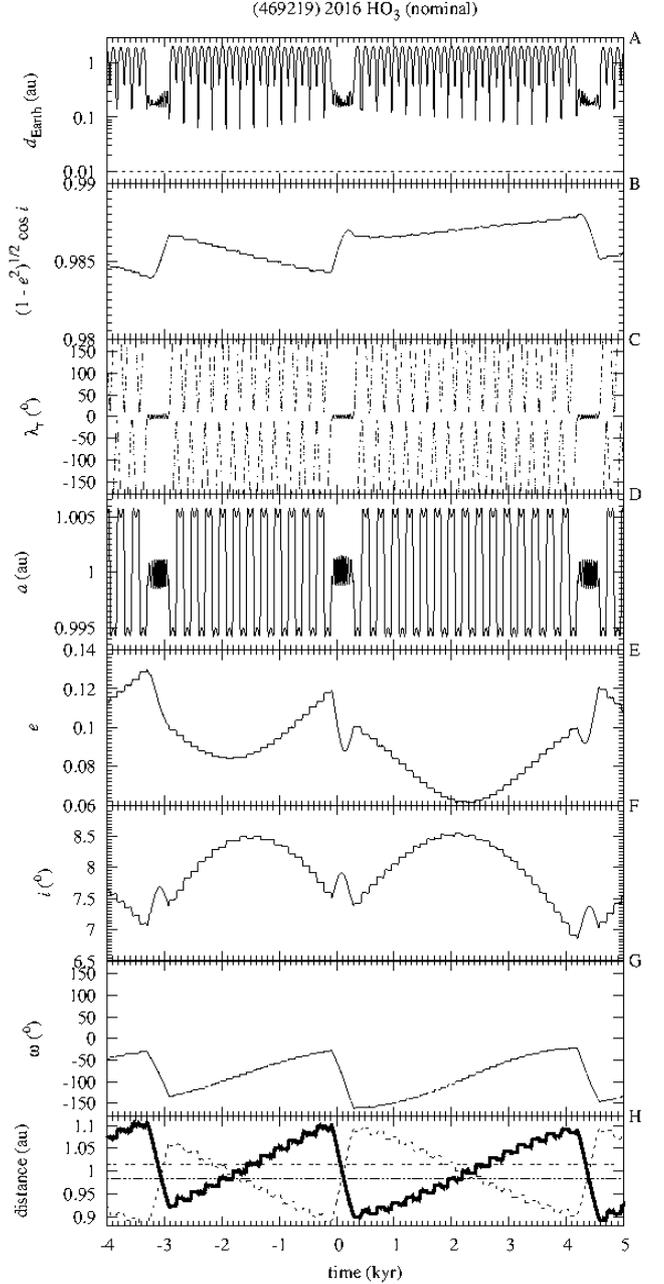}
        \caption{The various panels show the time evolution of relevant parameters for the nominal orbit of (469219) 2016~HO$_{3}$ in Table 
                 \ref{elements} during the time interval ($-$4, 5) kyr. Panel A shows how the distance from the Earth changes over time and 
                 the geocentric distance equivalent to the value of the radius of the Hill sphere of the Earth, 0.0098 au, is plotted as a 
                 dashed line. Panel B displays the evolution of the Kozai-Lidov parameter $\sqrt{1 - e^2} \cos i$. Panel C shows how the 
                 value of the resonant angle, $\lambda_{\rm r}$, changes during the displayed time interval. Panel D displays the behaviour
                 of the semimajor axis, $a$. Panel E shows how the orbital eccentricity, $e$, changes over time. Panel F plots the value of 
                 the orbital inclination, $i$. Panel G displays how the value of the argument of perihelion, $\omega$, changes over time.
                 Panel H shows the values of the distances to the descending (thick line) and ascending nodes (dotted line) and those of 
                 Earth's aphelion and perihelion distances. The nodal distances have been computed using Equation (\ref{nodeseq}). 
                }
        \label{short}
     \end{figure}
%
%

     \subsection{Recurrent co-orbital dynamics}
        Fig. \ref{short} shows the time evolution of various parameters for the nominal orbit of (469219) 2016~HO$_{3}$ in Table \ref{elements} 
        during the time interval ($-$4, 5) kyr. The evolution of the critical angle around $t=0$ displayed in panel C confirms that the 
        value of $\lambda_{\rm r}$ librates about 0\degr, which is the condition for being engaged in quasi-satellite behaviour. Panel C 
        also shows that 469219 switches repeatedly between the quasi-satellite and horseshoe configurations. Its current quasi-satellite 
        episode started nearly 100 yr ago and it will end in about 300 yr from now. The object experiences three quasi-satellite episodes 
        and their associated transitions to and from the horseshoe librator dynamical state during the displayed timeframe. These 
        transitions take place at quasi-regular intervals as seen in panel C. During the quasi-satellite episodes, the value of $e$ is the 
        highest (see panel E) and that of $i$, the lowest (see panel F); the value of the argument of perihelion decreases (see panel G) as 
        predicted by Namouni (1999). 

\begin{figure*}
       \centering
        \includegraphics[width=\linewidth]{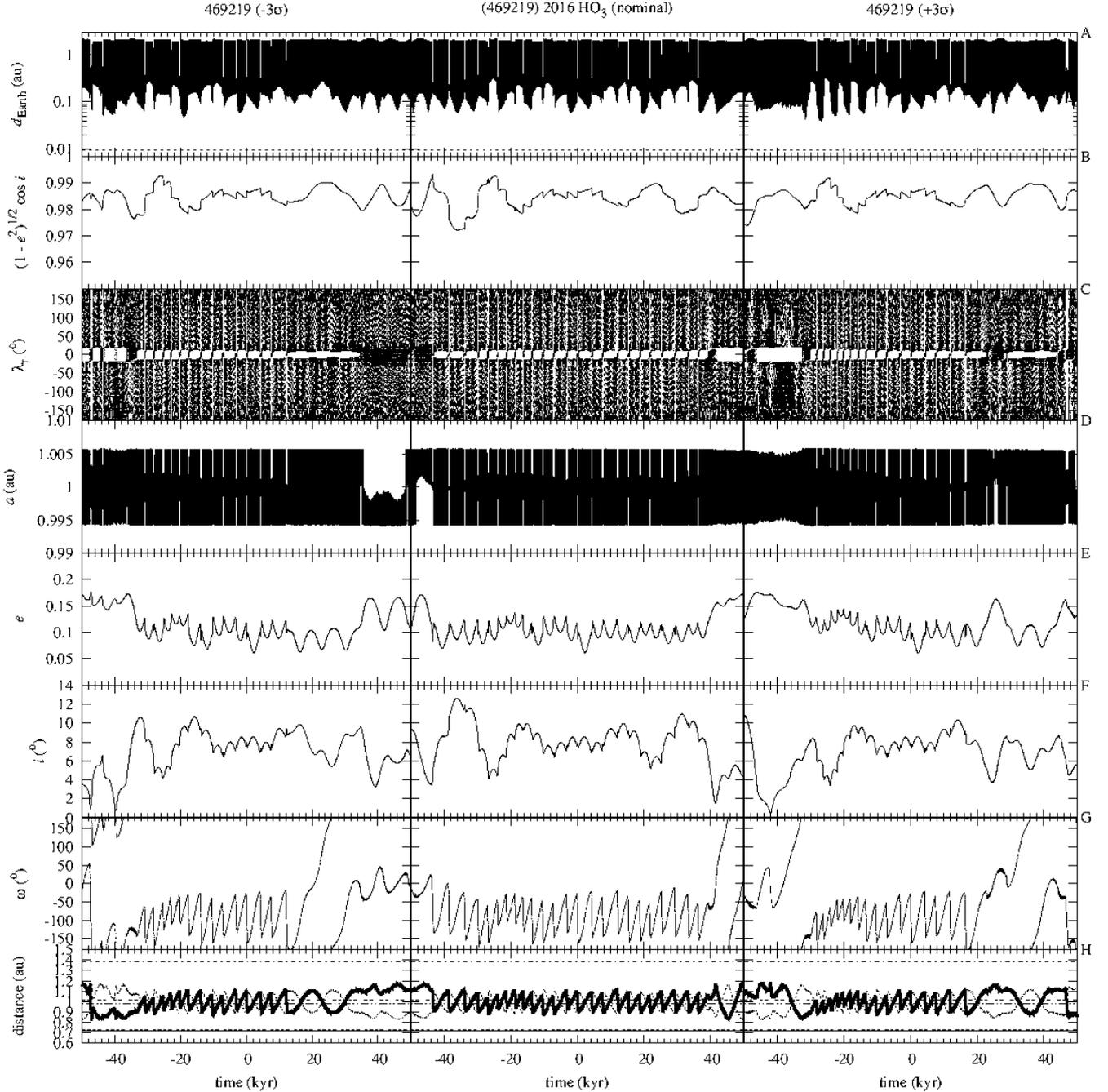}
        \caption{As Fig. \ref{short} but displaying the entire integration for the nominal orbit of (469219) 2016~HO$_{3}$ as in Table 
                 \ref{elements} (central panels) and two representative examples of orbits that are quite different from the nominal one 
                 (see the text for details). In the H-panels, Earth's, Venus' and Mars' aphelion and perihelion distances are also shown.
                }
        \label{control}
     \end{figure*}

        The orbital behaviour observed in Fig. \ref{short} is very similar to the one displayed by Earth co-orbital 2015~SO$_{2}$ in fig. 2 
        of de la Fuente Marcos \& de la Fuente Marcos (2016a). The short-term dynamical evolution of various parameters is virtually 
        identical for these two objects although they are out of phase by a few centuries and the average duration of the quasi-satellite 
        episodes of 469219 is longer, 400~yr versus 150~yr for 2015~SO$_{2}$. The mechanism that controls the transitions is also analogous 
        (see below). Fig. \ref{short} is a detailed view of the entire integration displayed in Fig. \ref{control}, central panels. The 
        behaviour of the various parameters is very similar along the entire simulated time; the evolution of the nominal orbit of 469219 is 
        unusually stable. About 24 (a maximum of 35 has been found for other control orbits) quasi-satellite episodes take place in 100 kyr 
        of simulated time. Among similarly behaved co-orbitals like 2015~SO$_{2}$, the recurrent co-orbital dynamics displayed by 469219 is 
        the lengthiest.  

        Asteroid 469219 becomes a quasi-satellite of the Earth whenever its descending node is farthest from the Sun and its ascending node 
        is closest to the Sun (see Fig. \ref{short}, panel H). The object becomes a horseshoe when the nodal positions are inverted with 
        respect to the previous situation (see Fig. \ref{short}, panel H). We pointed out above that the current value of the MOID of this 
        object is 0.0345~au, which is not particularly small when compared with the value of the radius of the Hill sphere of the Earth, 
        0.0098 au. Therefore, no truly close approaches are expected and this is what is seen in Fig. \ref{control}, panel A; this behaviour 
        has been observed for all the integrated control orbits. Close encounters with the Earth--Moon system are only possible in the 
        vicinity of the nodes. For a prograde orbit, the distance between the Sun and the nodes is given by the expression:
        \begin{equation}
           r=a(1 - e^2)/(1\pm{e}\cos\omega)\,, \label{nodeseq}
        \end{equation}
        where the "+" sign is for the ascending node (where the orbit crosses the Ecliptic from South to North) and the "$-$" sign is for 
        the descending node. Both distances appear in Figs \ref{short} and \ref{control}, panel H. Fig. \ref{short}, panels C and H, shows 
        that transitions between co-orbital states occur when the nodes of the orbit of this minor body are farthest from the Earth and the 
        average gravitational influence of the Earth--Moon system on the asteroid is the weakest. Therefore, at those moments, the secular 
        influence of more distant planets must be at its peak. 

        Jupiter is the dominant secular perturber for this object (see the discussion in the following section). During the quasi-satellite 
        phase both nodes cross the path of the Earth; then, distant encounters with the Earth--Moon system are possible at both nodes (see
        Fig. \ref{short}, panel H) and the sustained action of these encounters slowly increases the asteroid's orbital energy making the 
        transition to a horseshoe trajectory possible. These relatively distant close approaches are well beyond the radius of the Hill 
        sphere (see Fig. \ref{control}, A-panels) and cannot perturb the orbit dramatically. The described mechanism (see also de la Fuente 
        Marcos \& de la Fuente Marcos 2016a) is also affecting other Earth co-orbitals. In addition to 2015~SO$_{2}$ (and now 469219), 
        asteroids 2001~GO$_{2}$, 2002~AA$_{29}$ and 2003~YN$_{107}$ are well-documented cases of such a dynamical behaviour (Brasser et al. 
        2004). The existence of these transitions had been previously predicted and explained by Namouni (1999) and Christou (2000a). 
%
%
%
%

     \subsection{Kozai-Lidov resonance? not quite}
        Fig. \ref{short} shows coupled oscillations in eccentricity (panel E), inclination (panel F) and argument of perihelion (panel G). 
        These are widely considered as the signposts of the Kozai-Lidov mechanism (Kozai 1962; Lidov 1962); also, the value of the 
        Kozai-Lidov parameter (panel B) remains fairly constant. In our case and in principle, the Kozai-Lidov scenario is defined by the 
        presence of a primary (the Sun in our case), a perturbed body (an asteroid), and a massive outer perturber (Jupiter in our case) 
        such as the ratio of semimajor axes (perturbed versus perturber) tends to zero. Under this orbital architecture the libration of the 
        argument of perihelion occurs at $\omega$ = 90\degr or 270\degr. This is observed in Fig. \ref{short}, panel G, where $\omega$ 
        librates about 270\degr (or $-$90\degr). Under these conditions, aphelion always occurs away from the orbital plane of the 
        perturber. A classical example of a minor body subjected to the Kozai-Lidov effect induced by an outer perturber is the asteroid 
        (3040) Kozai that is perturbed by Jupiter. 

        Another variant of the Kozai-Lidov scenario is found when the ratio of semimajor axes (perturbed versus perturber) is close to one 
        which is typical of asteroids trapped in a 1:1 mean motion resonance with a host planet. In this case, the libration occurs at 
        argument of perihelion equal to 0\degr or 180\degr. In this alternative scenario, the nodes are located at perihelion and at 
        aphelion, i.e. away from the massive perturber (see e.g. Milani et al. 1989; Michel \& Thomas 1996). If $\omega$ librates about 
        0\degr then the object subjected to the Kozai-Lidov secular resonance reaches perihelion at nearly the same time it crosses the 
        ecliptic plane from South to North (ascending node); conversely, if $\omega$ oscillates about 180\degr then the Kozai-Lidov librator 
        reaches perihelion while approaching the descending node. Objects trapped at either version of the Kozai-Lidov resonance 
        ($\omega\sim90\degr$, 270\degr or $\omega\sim0\degr$, 180\degr) exhibit a precession rate of $\omega$ compatible with zero. 

\begin{figure}
        \centering
         \includegraphics[width=\linewidth]{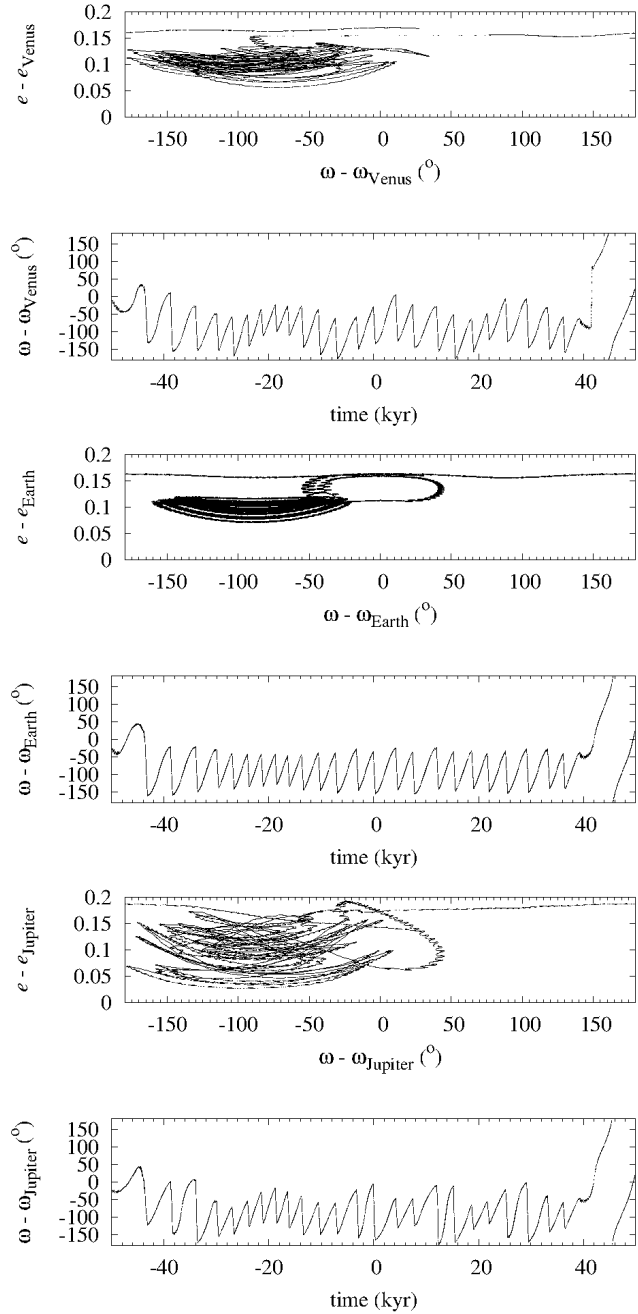}
         \caption{The $e_{\rm r} \omega_{\rm r}$-portrait relative to Venus, the Earth, and Jupiter for (469219) 2016~HO$_{3}$.
                 }
         \label{fEWfc71M}
      \end{figure}

\begin{figure}
        \centering
         \includegraphics[width=\linewidth]{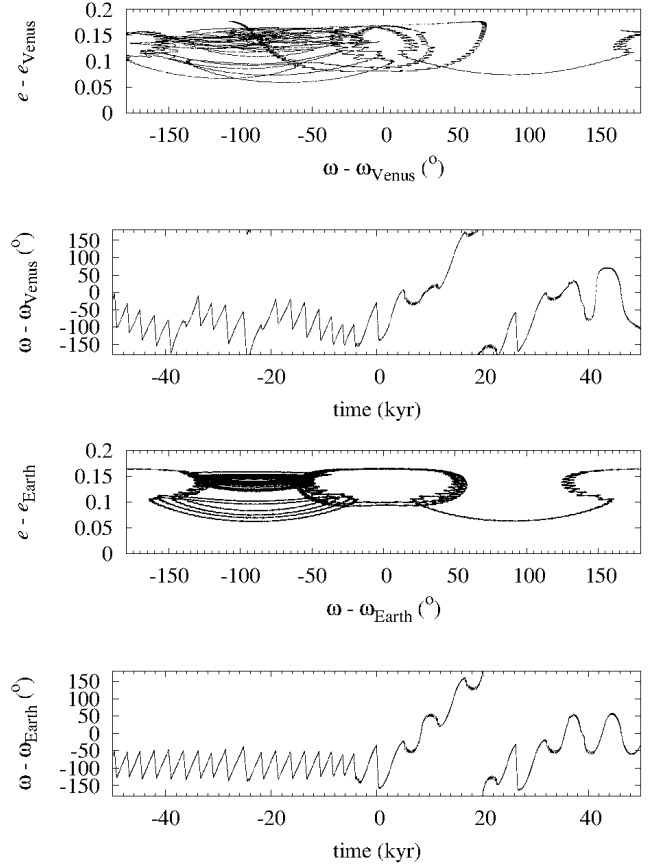}
         \caption{The $e_{\rm r} \omega_{\rm r}$-portrait relative to Venus and the Earth for (469219) 2016~HO$_{3}$, if Jupiter is removed
                  from the calculations.
                 }
         \label{fEWfc71Mwj}
      \end{figure}

        In principle, Fig. \ref{control}, central panel G, shows that (469219) 2016~HO$_{3}$ may have been locked in a Kozai-Lidov resonance 
        with $\omega$ librating about 270\degr for nearly 100 kyr and probably more. Because of the Kozai-Lidov resonance, both $e$ (central 
        panel E) and $i$ (central panel F) oscillate with the same frequency but out of phase (for a more detailed view, see Fig. \ref{short}, 
        panels E and F); when the value of $e$ reaches its maximum the value of $i$ is the lowest and vice versa ($\sqrt{1 - e^2} \cos i \sim$ 
        constant, see Fig. \ref{short}, panel B). During the simulated time and for the nominal orbit, 469219 reaches perihelion and 
        aphelion the farthest possible from the ecliptic. Fig. \ref{control}, G-panels, show that for other incarnations of the orbit of 
        469219, different from the nominal one, $\omega$ may librate about 90\degr as well during the simulated time interval. However, is 
        this a true Kozai-Lidov resonance? Namouni (1999) has shown that the secular evolution of co-orbital objects is viewed more 
        naturally in the $e_{\rm r} \omega_{\rm r}$-plane, where $e_{\rm r}$ and $\omega_{\rm r}$ are the relative eccentricity and argument 
        of perihelion computed as defined in Namouni's work (see equations 3 in Namouni 1999); these are based on the vector eccentricity 
        and the vector inclination. Fig. \ref{fEWfc71M} shows the multi-planet $e_{\rm r} \omega_{\rm r}$-portrait for the nominal orbit of 
        this object. It clearly resembles figs 13 and 19 in Namouni (1999). Asteroid 469219 librates around $\omega_{\rm r}=-90\degr$ for 
        Venus, the Earth, and Jupiter. This behaviour corresponds to domain III in Namouni (1999), horseshoe-retrograde satellite orbit 
        transitions and librations (around $\omega_{\rm r}=-90\degr$ or 90\degr). For a given cycle, the lower part corresponds to the 
        horseshoe phase and the upper part to the quasi-satellite or retrograde satellite phase. This is not the Kozai-Lidov resonance; in 
        this case, the Kozai-Lidov domain (domain II in Namouni 1999) is characterized by libration around $\omega_{\rm r}=0\degr$ (or 
        180\degr) which is only briefly observed at the end of the backwards integrations (see Fig. \ref{fEWfc71M}). The Kozai-Lidov 
        resonance is however in action at some stage in the orbits displayed in Figs \ref{control} and \ref{control2}. Our calculations show 
        that the orbital evolution followed by 469219 is the result of the dominant secular perturbation of Jupiter as the periodic 
        switching between co-orbital states ceases after about 8 kyr if Jupiter is removed from the calculations. Fig. \ref{fEWfc71Mwj} 
        shows that, without Jupiter, 469219 switches between the Kozai-Lidov domain and that of horseshoe-quasi-satellite orbit transitions 
        and librations (including both $-$90\degr and 90\degr). Jupiter plays a stabilizing role in the dynamics of objects following orbits 
        similar to that of 469219. It is not surprising that Jupiter instead of the Earth or Venus is acting as main secular perturber of 
        469219. Ito \& Tanikawa (1999) have shown that the inner planets share the effect of the secular perturbation from Jupiter; in fact, 
        Venus and our planet exchange angular momentum (Ito \& Tanikawa 2002). In their work, these authors argue that the inner planets 
        maintain their stability by sharing and weakening the secular perturbation from Jupiter. Tanikawa \& Ito (2007) have extended this 
        analysis to conclude that, regarding the secular perturbation from Jupiter, the terrestrial planets form a collection of loosely 
        connected mutually dynamically dependent massive objects. The existence of such planetary grouping has direct implications on the 
        dynamical situation studied here; if Jupiter is removed from the calculations, the overlapping secular resonances and the recurrent 
        dynamics disappear as well.
%
%
%
%
%
%
%
%

     \subsection{The impact of errors}
        In contrast with other Earth co-orbitals or Kozai librators (see e.g. de la Fuente Marcos \& de la Fuente Marcos 2015c, 2016b), 
        (469219) 2016~HO$_{3}$ follows a rather stable orbit that is only subjected to the direct perturbation of the Earth--Moon system. 
        Its current orbital solution (see Table \ref{elements}) has very small associated uncertainties, but moving embedded in a web of 
        secular resonances and switching regularly between co-orbital states, can induce unexpected instabilities.

\begin{figure*}
       \centering
        \includegraphics[width=\linewidth]{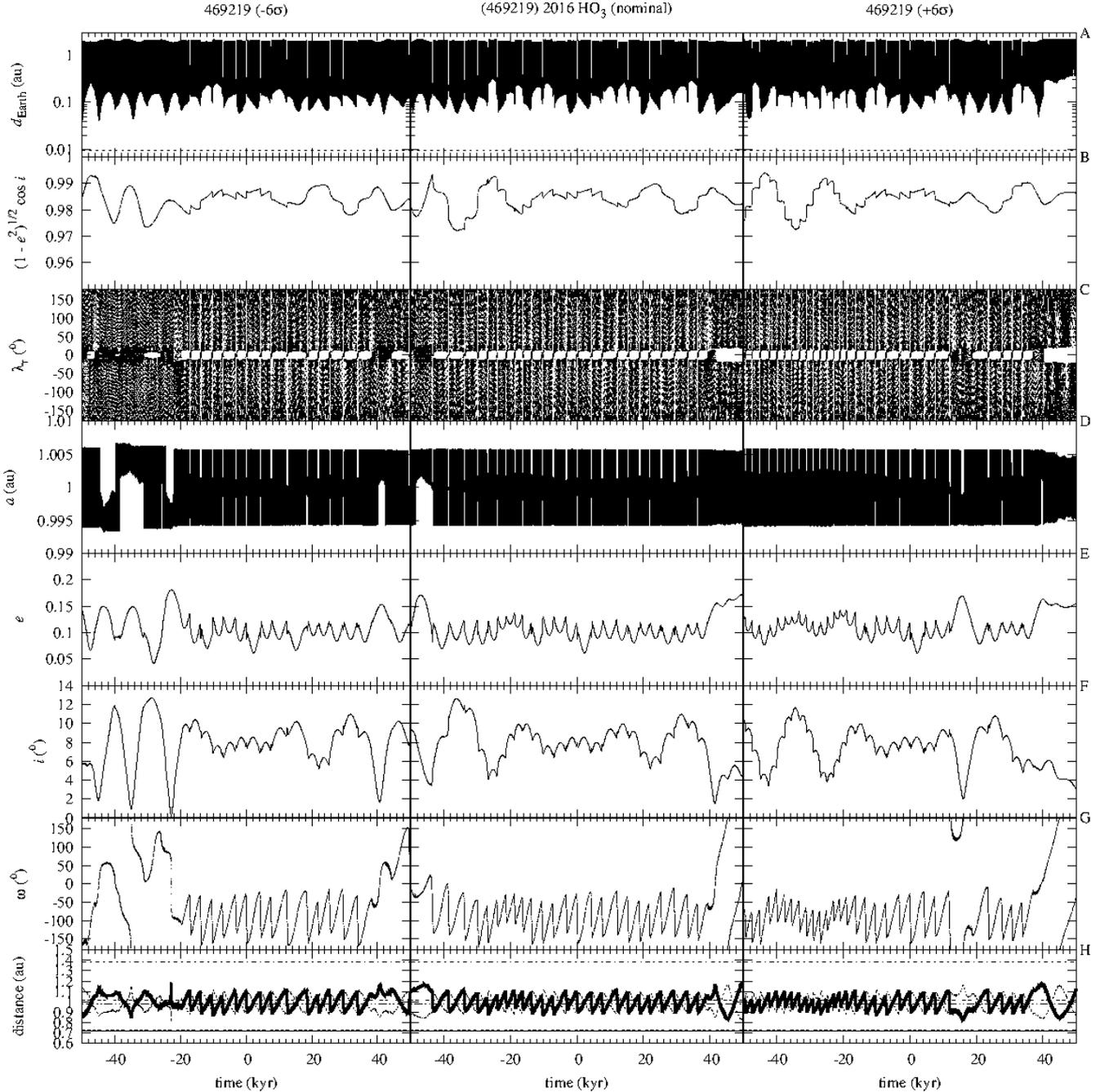}
        \caption{As Fig. \ref{control} but for two very different orbits (see the text for details). 
                }
        \label{control2}
     \end{figure*}

        In addition to the integrations that make use of the nominal orbital elements in Table \ref{elements} as initial conditions, we 
        have performed 50 control simulations with series of orbital parameters drawn from the nominal ones within the quoted uncertainties 
        and assuming Gaussian distributions for them. In the computation of these additional sets of orbital elements, the Box-Muller method 
        (Box \& Muller 1958; Press et al. 2007) has been applied to produce random numbers according to the standard normal distribution with 
        mean 0 and standard deviation 1. When computers are used to generate a uniform random variable ---in our case to seed the Box-Muller 
        method--- it will inevitably have some inaccuracies because, numerically, there is a lower bound on how close numbers can be to 0. 
        For a 64 bits computer the smallest non-zero number is $2^{-64}$ which means that the Box-Muller method will not produce random 
        variables more than 9.42 standard deviations from the mean (Press et al. 2007). Representative results from these calculations are 
        displayed in Figs \ref{control} and \ref{control2}. Here, when an orbit is labelled `$\pm3\sigma$' (Fig. \ref{control}), it has been 
        obtained by adding (+) or subtracting ($-$) three times the uncertainty from the orbital parameters (the six elements) in Table 
        \ref{elements}; an equivalent approach has been followed for orbits labelled `$\pm6\sigma$' (Fig. \ref{control2}).
%
%
%
%

        Figs \ref{control} and \ref{control2} show that the orbital evolution of this object is similar within $\pm$15~kyr of $t=0$. Beyond 
        this time interval and according to the results of the multiplicity of simulations performed, the object may start being subjected 
        to the Kozai-Lidov resonance but it will remain confined inside Earth's co-orbital region, in particular, evolving dynamically as a 
        horseshoe librator. This object never approaches the Earth under 0.04~au within $\pm$50~kyr of $t=0$. Recurrent transitions between 
        the horseshoe and quasi-satellite dynamical states are observed for all the simulated control orbits.

\begin{figure*}
      \centering
       \includegraphics[width=0.33\linewidth]{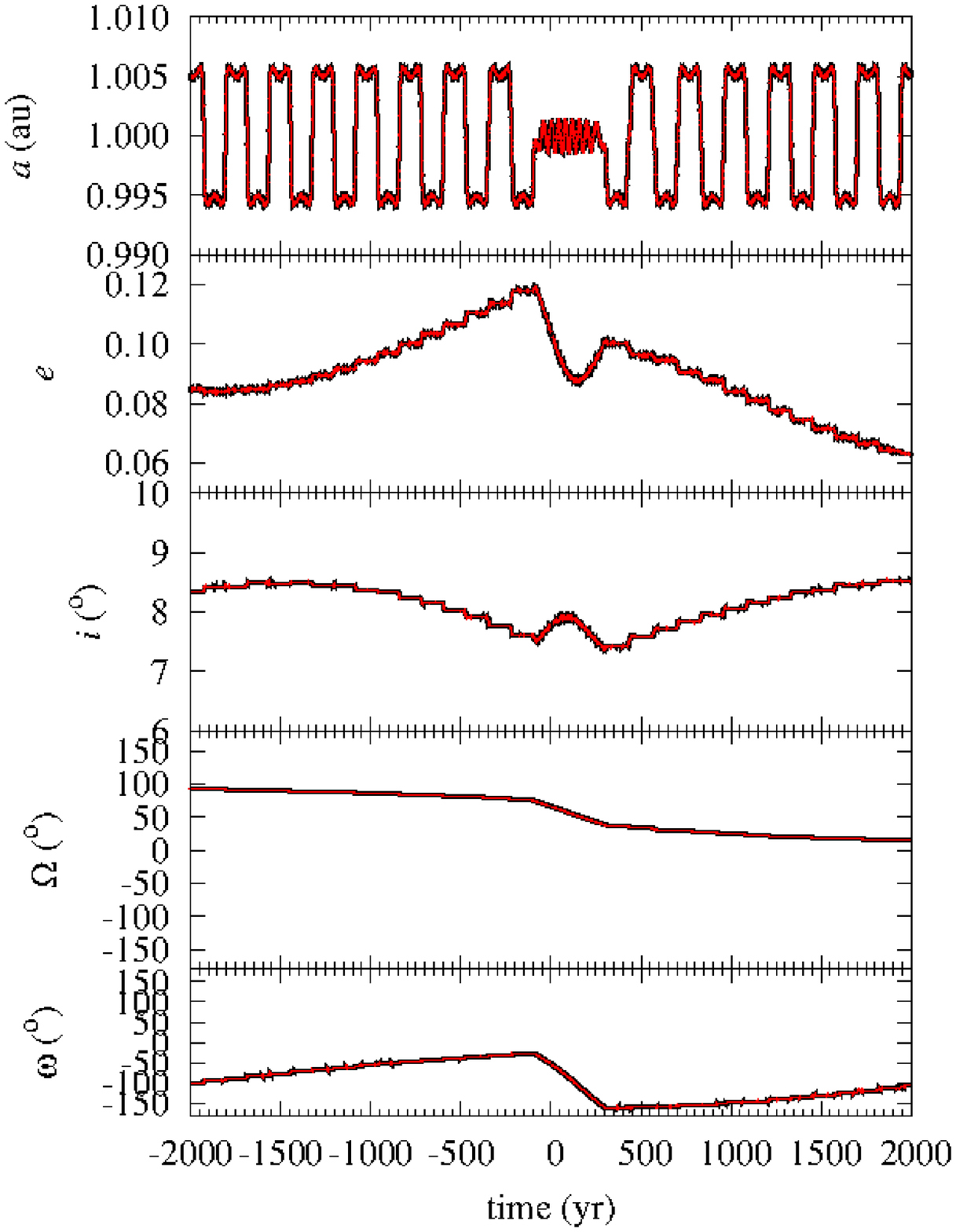}
       \includegraphics[width=0.33\linewidth]{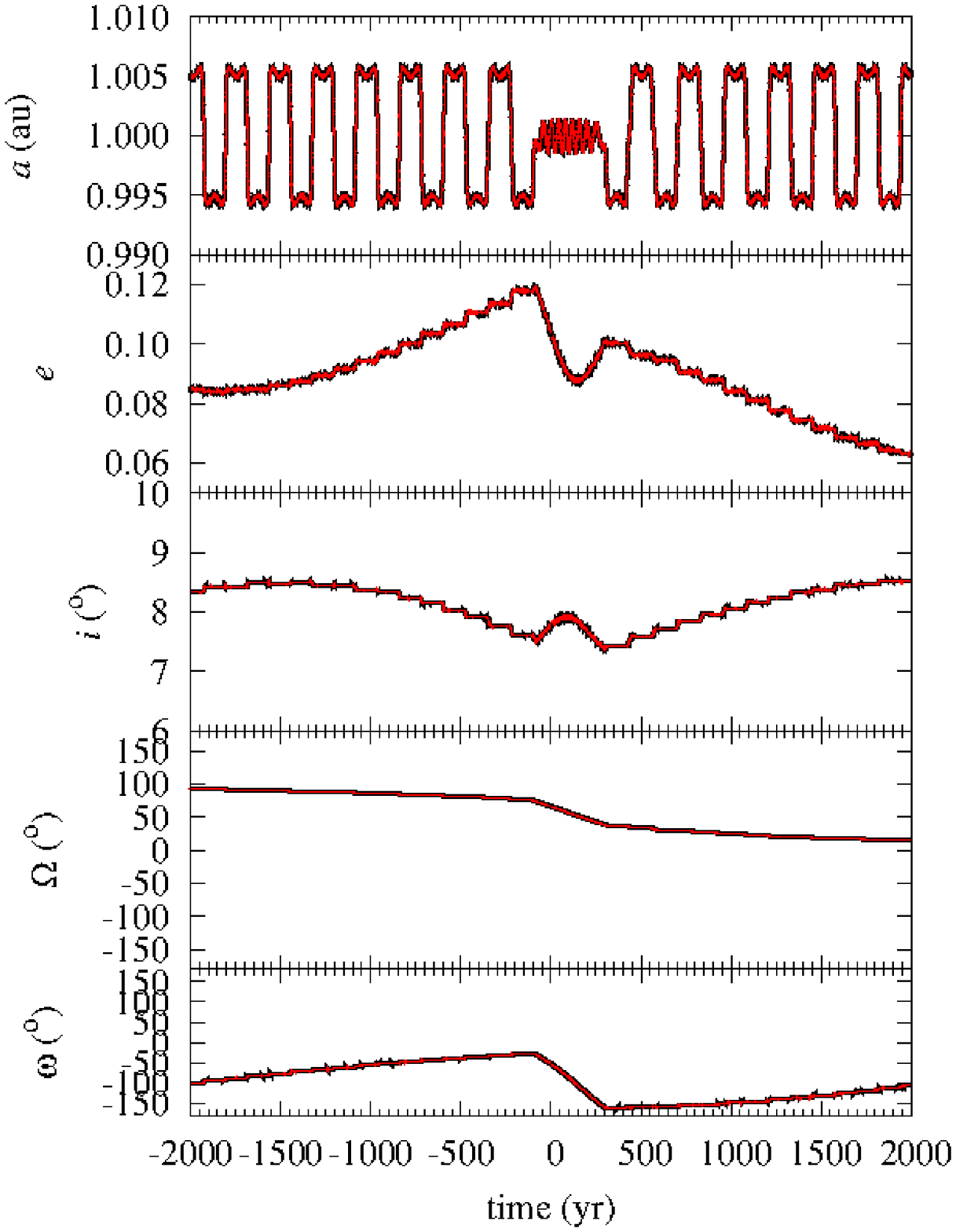}
       \includegraphics[width=0.33\linewidth]{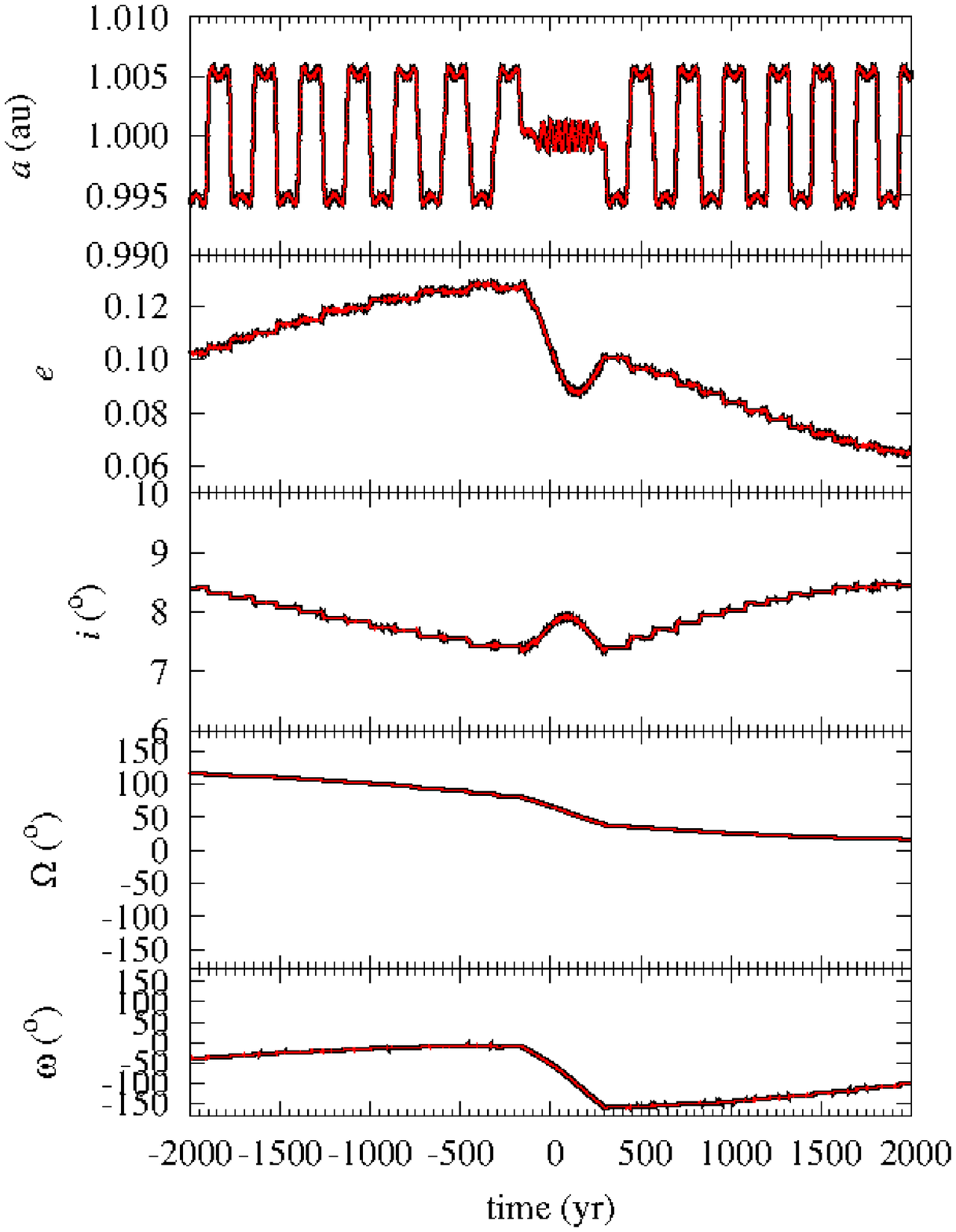}
       \caption{Time evolution of the orbital elements of (469219) 2016~HO$_{3}$. The black thick curve displays the average evolution of 
                100 control orbits, the red thin curves show the ranges in the values of the orbital parameters at the given time. Results 
                for a 1$\sigma$ (left-hand panels) and a 6$\sigma$ (central panels) spread in the initial values of the orbital elements 
                (classical approach, see the text for details), and using MCCM (see the text, right-hand panels).
               }
       \label{errors}
    \end{figure*}

        A more rigorous and detailed evaluation of the role of errors on our short-term results has been carried out by studying how the 
        changing values of the orbital parameters of the test orbits at $t=0$ influence the evolution of the osculating orbital elements as
        the simulation progresses. Three additional sets of 100 shorter control simulations ($\pm$2~kyr of $t=0$) are analysed here. The 
        first set (see Fig. \ref{errors}, left-hand panels), has been generated as previously described with the initial orbital elements of 
        each control orbit varying randomly, within the ranges defined by their mean values and standard deviations. For instance, new 
        values of the semimajor axis have been found using the expression $a_{\rm t} = \langle{a}\rangle + n \ \sigma_{a}\,r_{\rm i}$, where 
        $a_{\rm t}$ is the semimajor axis of the control orbit, $\langle{a}\rangle$ is the value of the semimajor axis from the nominal 
        orbit (Table \ref{elements}), $n$ is an integer (in our case 1, left-hand panels, or 6, central panels), $\sigma_{a}$ is the 
        standard deviation of $a$ supplied with the nominal orbit (Table \ref{elements}), and $r_{\rm i}$ is a (pseudo) random number with 
        standard normal distribution (see above). For this object, the errors are so small and the orbital evolution so smooth that nominal 
        and error-based results fully overlap within $\pm$2~kyr of $t=0$. Artificially increasing the values of the standard deviations 
        ---up to 6 times in the simulations displayed in Fig. \ref{errors}, right-hand panels--- gives nearly the same results. 
%
%
%
%

        Sitarski (1998, 1999, 2006) has shown that the somewhat classical approach used above is equivalent to considering a set of distinct 
        virtual minor bodies following similar orbits, but not a sample of test orbits resulting from an actual set of observations 
        associated with a single object. The orbit in Table \ref{elements} matches all the available observations for 469219 within certain, 
        very strict tolerances. In this context, the statistically correct procedure to compute control orbits is to consider how the 
        elements influence each other and their associated uncertainties, applying the Monte Carlo using the Covariance Matrix (MCCM) 
        approach (Bordovitsyna, Avdyushev \& Chernitsov 2001; Avdyushev \& Banschikova 2007). 

        Fig. \ref{errors}, right-hand panels, shows the results of a third set of short simulations whose initial conditions have been 
        generated using the implementation of the MCCM approach discussed in de la Fuente Marcos \& de la Fuente Marcos (2015b). The control 
        orbits studied here have initial parameters drawn from the nominal orbit (Table \ref{elements}) adding random noise on each initial 
        orbital element as described by the covariance matrix. The covariance matrix used here was provided by the JPL Small-Body 
        Database\footnote{http://ssd.jpl.nasa.gov/sbdb.cgi} and it has been computed at epoch 2457232.5 TDB. Our results show that, for this 
        particular object, the MCCM and the classical approaches produce consistent results; this is often the case when precise orbital 
        solutions of stable orbits are used (see e.g. de la Fuente Marcos \& de la Fuente Marcos 2015b, 2016a). We can confirm that 
        469219 is a present-day quasi-satellite of our planet and the probability of this assessment being incorrect is virtually 
        zero as our results are based on the analysis of nearly 2000 control orbits.

     \subsection{Long-term stability}
        The stability of quasi-satellite orbits has been explored in the theoretical works of Mikkola et al. (2006), Sidorenko et al. (2014) 
        and Pousse et al. (2016). Arguably, the most stable known Earth's co-orbital is 2010~SO$_{16}$. This object stays as horseshoe 
        librator for at least 120 kyr and possibly for up to 1 Myr (Christou \& Asher 2011). This is quite remarkable because it remains in 
        the same co-orbital state during this time span with no transitions. 

\begin{figure}
       \centering
        \includegraphics[width=\linewidth]{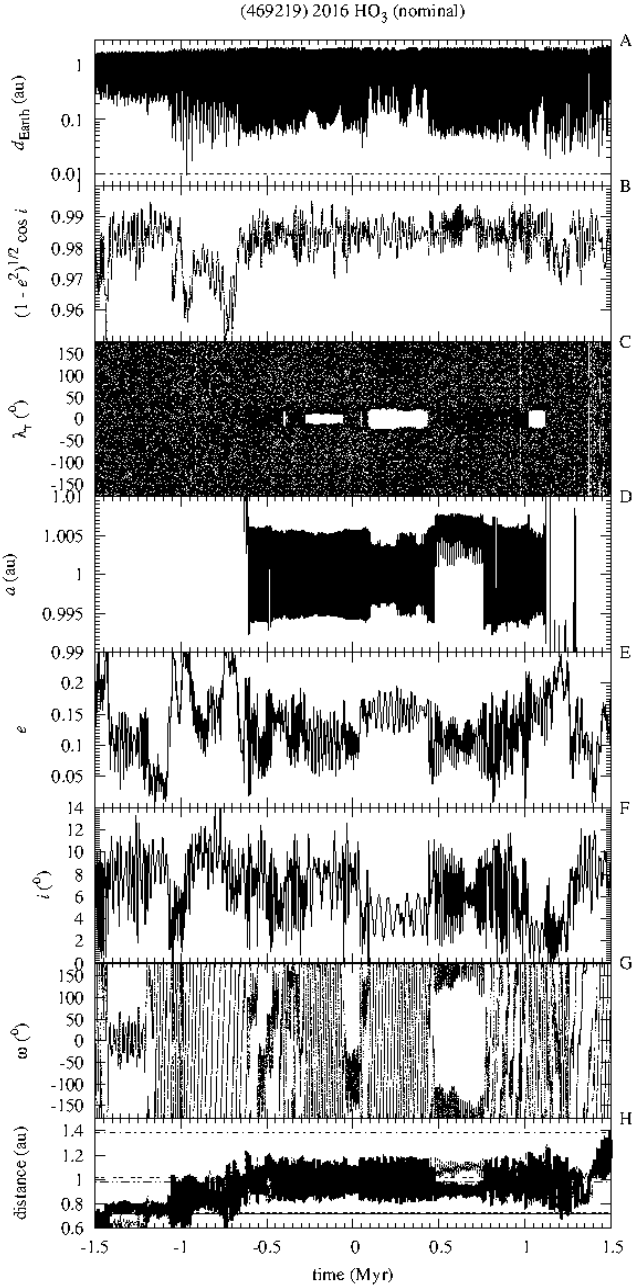}
        \caption{As in Fig. \ref{short} but displaying an extended integration.
                }
        \label{stability}
     \end{figure}

        Fig. \ref{stability} is an extension of the data displayed in Figs \ref{control} and \ref{control2}, central panels and it shows the 
        long-term dynamical evolution of the nominal orbit of (469219) 2016~HO$_{3}$ (Table \ref{elements}). Asteroid 469219 could be as 
        stable as 2010~SO$_{16}$, perhaps even more stable, but it experiences transitions between co-orbital states. However, if we assess 
        the overall stability of an Earth co-orbital in terms of how much time the object stays confined within Earth's co-orbital zone 
        ---that, based on the results of the integrations performed, currently goes from $\sim$0.994~au to $\sim$1.006~au--- then the object 
        discussed here is, with little doubt, the most stable known Earth co-orbital. A quantitative measure of the level of dynamical 
        stability associated with the orbital solution currently available for 469219 is in the value of its Lyapunov time ---or time-scale 
        for exponential divergence of integrated orbits starting arbitrarily close to each other--- that is $\sim$7500~yr; this time-scale 
        is nearly the same for both forward and backwards integrations. 
%
%
%
%

     \subsection{Objects in similar orbits}
        Minor bodies as small as (469219) 2016~HO$_{3}$ are not expected to be primordial, but fragments of larger objects. In this case, it 
        is logical to assume that other small bodies could be moving along paths similar to that of 469219 if they are trapped in some web 
        of secular resonances like the one described in the previous sections. In order to test this reasonable hypothesis, here we use the 
        D-criteria of Southworth \& Hawkins (1963), $D_{\rm SH}$, Lindblad \& Southworth (1971), $D_{\rm LS}$ (in the form of equation 1 in 
        Lindblad 1994 or equation 1 in Foglia \& Masi 2004), Drummond (1981), $D_{\rm D}$, and the $D_{\rm R}$ from Valsecchi, Jopek \& 
        Froeschl\'e (1999) to search for a possible dynamical link between 469219 and other known minor bodies. 

        An exploration of all the NEOs currently catalogued (as of 2016 July 7) by the JPL Small-Body Database$^9$ using these criteria 
        produced the list in Table \ref{similar}. In this list, objects are sorted by ascending $D_{\rm LS}$ and only those with $D_{\rm LS}$ 
        and $D_{\rm R} < 0.05$ are shown. The list includes 2015 SO$_{2}$ that is a previously documented horseshoe librator (de la Fuente 
        Marcos \& de la Fuente Marcos 2016a). It has already been pointed out that the orbital evolution of 2015 SO$_{2}$ resembles, in many 
        aspects, that of 469219 but, without proper integrations, it is not possible to draw any conclusions about the other objects 
        in Table \ref{similar}. Given the high degree of dynamical chaos that characterizes the neighbourhood of the orbit of our planet,
        finding two or more similar present-day orbits is not enough to claim a relationship, dynamical or otherwise, among them; a 
        representative set of test orbits must be integrated to show that the dynamical evolution of the candidate related objects over a 
        reasonable time interval is also similar (see e.g. Jopek \& Williams 2013). 

\begin{figure*}
       \centering
        \includegraphics[width=\linewidth]{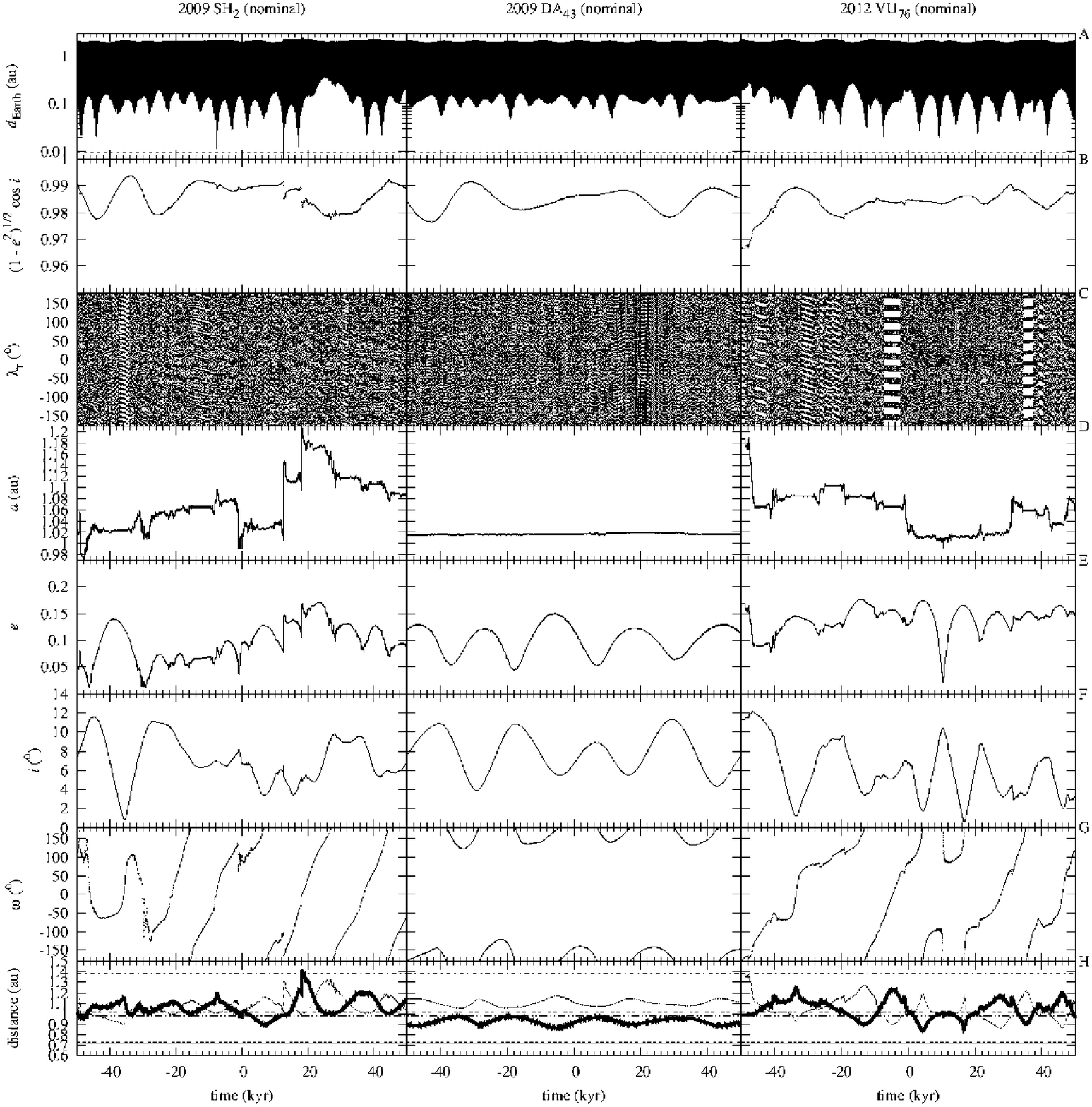}
        \caption{Analogous to Fig. \ref{control} for the nominal orbits of 2009 SH$_{2}$, 2009 DA$_{43}$ and 2012 VU$_{76}$.
                }
        \label{likes}
     \end{figure*}

        Fig. \ref{likes} shows the comparative orbital evolution of 2009 SH$_{2}$, 2009 DA$_{43}$ and 2012 VU$_{76}$. With the notable 
        exception of 2009 DA$_{43}$, the orbital evolution of the other two objects is rather unstable as expected from the low values of 
        their respective MOIDs (see Table \ref{similar}). None of the three objects is a present-day Earth co-orbital and they are located 
        well outside Earth's co-orbital zone. Asteroid 2009 DA$_{43}$ shows a very peculiar and stable orbital evolution controlled by the 
        Kozai-Lidov resonance; in this case, the value of its $\omega$ librates about 180\degr, a variant of the Kozai-Lidov mechanism that 
        arises when the ratio of semimajor axes of both perturbed and perturber bodies is close to one. For 2009 DA$_{43}$ the dominant 
        secular perturbation is that of the nearly co-orbital perturber (Earth).
%
%
     \begin{table*}
      \centering
      \fontsize{8}{11pt}\selectfont
      \tabcolsep 0.07truecm
      \caption{Orbital elements, orbital periods ($P$), perihelia ($q = a \ (1 - e)$), aphelia ($Q = a \ (1 + e)$), number of
               observations ($n$), data-arc, absolute magnitudes ($H$) and MOID of minor bodies with orbits similar to that of (469219) 
               2016~HO$_{3}$ (see Table \ref{elements}). The various $D$-criteria ($D_{\rm SH}$, $D_{\rm LS}$, $D_{\rm D}$ and $D_{\rm R}$) 
               are also shown. The objects are sorted by ascending $D_{\rm LS}$ (equation 1 in Lindblad 1994 or equation 1 in Foglia \& Masi 
               2004). Only objects with $D_{\rm LS}$ and $D_{\rm R} < 0.05$ are shown. The orbits are referred to the Epoch 2457600.5 
               (2016-July-31.0) TDB. Data as of 2016 July 7.}
      \begin{tabular}{lllllllllllllllll}
       \hline
          Asteroid       & $a$ (au)  & $e$        & $i$ (\degr) & $\Omega$ (\degr) & $\omega$ (\degr) & $P$ (yr) & $q$ (au) & $Q$ (au)
                         & $n$ & arc (d) & $H$ (mag) & MOID (au)
                         & $D_{\rm SH}$ & $D_{\rm LS}$ & $D_{\rm D}$ & $D_{\rm R}$ \\
       \hline
         2009 SH$_{2}$   & 0.99141   & 0.09427    & 6.81139     &   6.69012        & 101.64642        &  0.99    & 0.8979   & 1.0849 
                         & 109 & 14      & 24.90     & 0.00036
                         & 0.1942       & 0.0195       & 0.0829      & 0.0299      \\
         2009 DA$_{43}$  & 1.01681   & 0.11836    & 6.70098     & 157.87571        & 215.44085        &  1.03    & 0.8965   & 1.1372 
                         & 31  & 21      & 24.60     & 0.08263
                         & 0.1813       & 0.0235       & 0.0861      & 0.0486      \\
         2015 SO$_{2}$   & 0.99897   & 0.10818    & 9.18631     & 182.93194        & 290.04719        &  1.00    & 0.8909   & 1.1070 
                         & 84  & 9       & 23.90     & 0.01919
                         & 0.2983       & 0.0257       & 0.1007      & 0.0281      \\
         2012 VU$_{76}$  & 1.01824   & 0.12836    & 6.75802     &  52.37181        &  88.18982        &  1.03    & 0.8875   & 1.1489 
                         & 47  & 596     & 25.70     & 0.00376
                         & 0.2127       & 0.0314       & 0.1331      & 0.0462      \\
         2016 CO$_{246}$ & 1.00084   & 0.12413    & 6.42284     & 137.15957        & 118.36021        &  1.00    & 0.8766   & 1.1251 
                         & 21  & 27      & 26.00     & 0.03787
                         & 0.2452       & 0.0370       & 0.1248      & 0.0037      \\
       \hline
      \end{tabular}
      \label{similar}
     \end{table*}
%
%
%
%
%
%

  \section{The role of the uncertainties in the masses of the Jovian planets}
\begin{figure*}
       \centering
        \includegraphics[width=\linewidth]{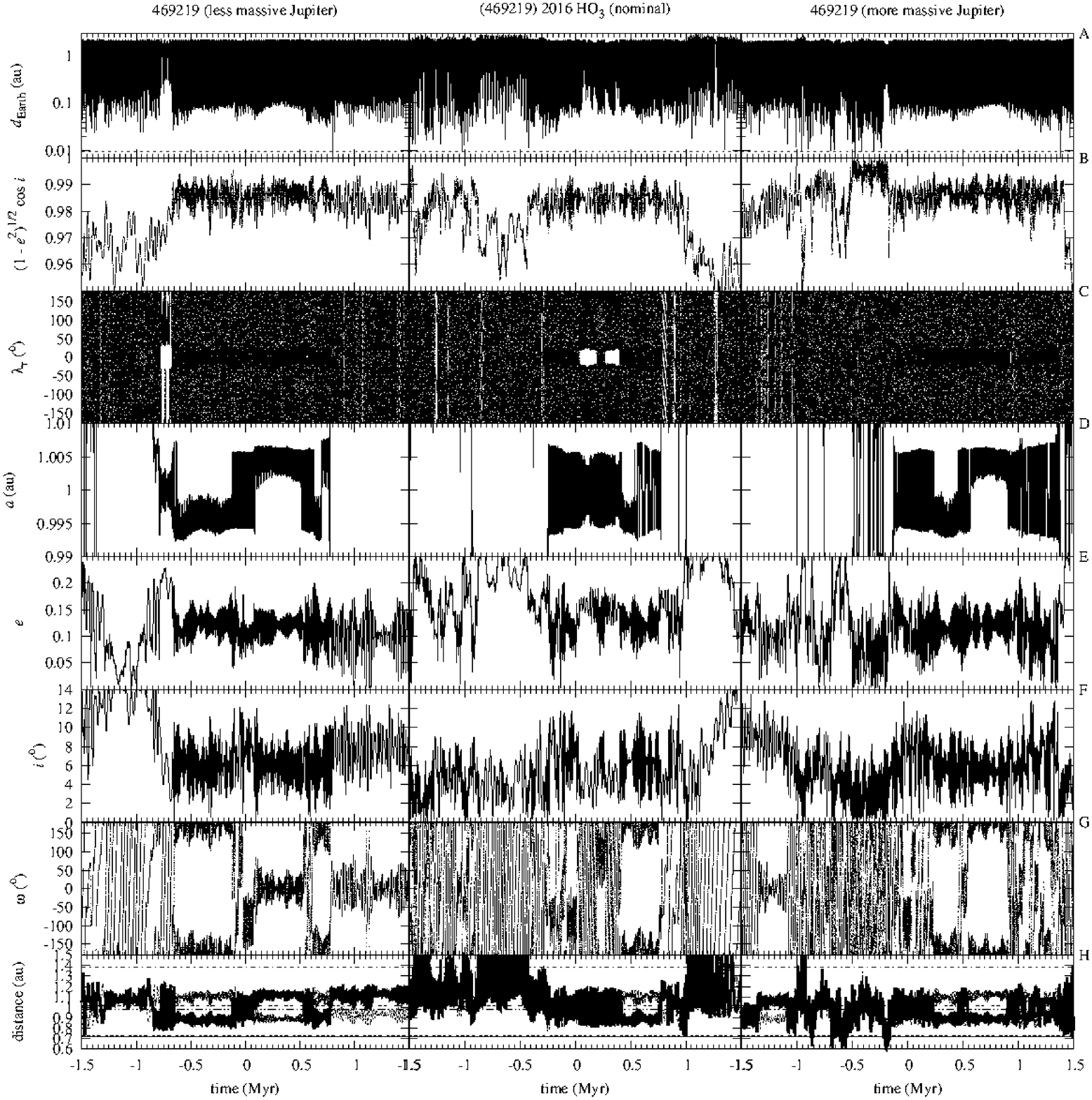}
        \caption{Comparative long-term dynamical evolution of various parameters (as in Fig. \ref{short}) for the nominal orbit of 
                 (469219) 2016~HO$_{3}$ as in Table \ref{elements} and three representative values of the mass of Jupiter, nominal 
                 value$-1\sigma$ (left-hand panels), nominal value (central panels), and nominal value$+1\sigma$ (JUP310 ephemeris, see the 
                 text for details).
                }
        \label{stability2}
     \end{figure*}

     Figs \ref{fEWfc71M} and \ref{fEWfc71Mwj} strongly suggest that Jupiter plays a major role on the secular evolution of (469219) 
     2016~HO$_{3}$. Our physical model assumes a nominal value for the mass of Jupiter that is affected by some uncertainty. An issue 
     frequently overlooked in the study of the orbital evolution of asteroids is the fact that the uncertainty in the values of the masses 
     of the Jovian planets is still significant. In the particular case of NEOs, many of them are embedded in a tangled web of secular 
     resonances. Such dynamical arrangement is particularly sensitive to the numerical values of the physical and orbital parameters of the 
     actors involved. Most dynamical studies focus exclusively on the role of the errors in the available orbital solutions of the various 
     bodies included in the calculations, but Jupiter is a major indirect NEO perturber (as it is in the case of 469219) and the value of 
     its mass is still in need of some improvement. The value of the mass of Jupiter as quoted by the JPL Solar System Dynamics Group, 
     Horizons On-Line Ephemeris System is $1.89813\pm0.00019\times10^{27}$~kg. Its uncertainty is about 217 times that of the value of the 
     mass of the Earth ($0.0006\times10^{24}$~kg) and it is nearly 0.022 $M_{\oplus}$ ($M_{\oplus}=5.97219\times10^{24}$~kg). This value of 
     the mass of Jupiter corresponds to the ephemeris JUP230\footnote{R. A. Jacobson, personal communication, Principal Engineer, Guidance, 
     Navigation, and Control Section, ``Jovian satellite ephemeris JUP230," Jet Propulsion Laboratory, Pasadena, California, 2005.} and it 
     has been used to obtain Fig. \ref{stability} and the other figures with the exception of Fig. \ref{stability2}. 

     If instead of using the ephemeris JUP230, we make use of the most updated ephemeris JUP310,\footnote{R. A. Jacobson, personal 
     communication, Principal Engineer, Guidance, Navigation, and Control Section, ``Jovian satellite ephemeris JUP310," Jet Propulsion 
     Laboratory, Pasadena, California, 2016.} the value of the mass of Jupiter is $1.89826\pm0.00023\times10^{27}$~kg. Fig. \ref{stability2} 
     shows the comparative long-term evolution of the values of various parameters (as in Fig. \ref{short}) for the nominal orbit of 469219 
     (Table \ref{elements}). Three representative values of the mass of Jupiter from the ephemeris JUP310 are considered, nominal value 
     minus 1$\sigma$ (left-hand panels), nominal value (central panels), and nominal value plus 1$\sigma$ (right-hand panels). Changing the 
     value of the mass of the main secular perturber has visible effects on the long-term orbital evolution of 469219. Small 
     differences already appear beyond $\pm$10 kyr of integrated time. In principle, a heavier Jupiter may further stabilize the dynamical 
     evolution of this object, but more calculations are needed to provide a statistically robust answer.

     Our results here should be understood as a cautionary note regarding long-term predictions of the orbital evolution of NEOs based on 
     the currently available values of the masses of the Jovian planets. It is unlikely that the masses of Uranus or Neptune will be 
     improved within the next decade or so, but the ongoing Juno mission (Bolton et al. 2010) should be able to reduce the degree of 
     uncertainty in the value of the mass of Jupiter considerably (Le Maistre et al. 2016). Such a strong improvement in the precision of 
     Jupiter's mass parameter determination will indirectly improve our assessment of the stability of Earth co-orbitals.
%
%
%
%

  \section{Asteroid (469219) 2016~HO$_{3}$: a suitable probe into the origins of Earth's co-orbital asteroid population}
     Among known Earth co-orbitals, the long-term evolution of (469219) 2016~HO$_{3}$ clearly stands out. The results of our calculations 
     strongly suggest that it may have been a companion to our planet for at least 1 Myr and perhaps more. Fig. \ref{stability} indicates 
     that this object may have arrived to Earth's co-orbital zone as early as 600 kyr ago, but probably even 1 Myr ago (not shown). It may 
     have been captured from the general NEO population but, being such a long-term companion to our planet, other scenarios regarding its 
     putative origin may also be plausible. Unlikely but not impossible is an artificial origin for this object. Albeit improbably, 469219 
     could be an artificial interloper, a relatively large piece of hardware from previous space missions returning to the neighbourhood of 
     the Earth--Moon system (see the discussion in section 9 of de la Fuente Marcos \& de la Fuente Marcos 2015c). In contrast with previous 
     sections, our discussion here is significantly more speculative because the influence of orbital chaos severely limits our chances of 
     determining with confidence the sources of the NEOs (Connors et al. 2004). 

     It is possible that some NEOs may have been produced in the Earth--Moon system (see e.g. Margot \& Nicholson 2003); however, this 
     scenario is disfavoured by other authors who argue that sources in the main asteroid belt are probably more likely (see e.g. Morais \& 
     Morbidelli 2002). The initial hypothesis suggesting an origin in the Earth--Moon system for some of these objects assumes that they 
     could be the result of impacts on the Moon (see e.g. Warren 1994; Gladman et al. 1995; Bottke et al. 1996; Gladman 1996); i.e., within 
     the framework of this hypothesis a fraction of NEOs are nothing but Lunar debris. However, an origin in the Earth--Moon system does not 
     require a physical impact on our satellite. 

     Objects like 469219 may be relatively recent fragments of an already co-orbital and older parent body. In the Earth's neighbourhood, at 
     least two processes can dislodge large rocks from a weakly bound minor body. Resurfacing events can be triggered when an asteroid 
     encounters our planet during flybys at relatively large planetary distances, in the range 5--20 planetary radii (see e.g. Keane \& 
     Matsuyama 2015); even more likely, fragments can also be released after the spin rate of an asteroid is dramatically altered during a 
     close encounter with our planet (Scheeres et al. 2005), such a sudden change may induce subsequent structural failure that results in 
     large rocky blocks being ejected from the larger, crumbling asteroid (see e.g. Denneau et al. 2015). In the framework of this 
     hypothetical ---but certainly not impossible--- scenario, the relatively large fragments released during one of these partial 
     disruption events may remain within the orbital neighbourhood of our planet confined inside a dynamical mesh where resonances 
     interweave in a complex pattern that pervades the entire region (see Section 3). 

     Due to the singular orbital evolution of this small body, it will remain observable from the ground under relatively favourable 
     conditions for many decades to come (see Figs \ref{qsradec}, \ref{qsradecperi} and \ref{short}). Its apparent visual magnitude at 
     perigee could be as low as 22~mag or slightly lower in April. These facts could turn this object into an eventual `Rosetta stone' whose 
     study could lead to uncovering the true origin of Earth's co-orbitals or at least of some of them. Spectroscopic observations during 
     its future close approaches to our planet should be able to confirm a possible genesis in the Earth--Moon system as Lunar debris or 
     even a physical connection with 2015 SO$_{2}$ which is already a close dynamical relative, and also discard a putative artificial 
     origin for this object. Such studies may be particularly helpful in improving our understanding of the origins ---local versus 
     captured--- of Earth's co-orbital asteroid population. 

  \section{Discussion}
     The subject of NEOs currently engaged in quasi-satellite behaviour with our planet has been revisited in de la Fuente Marcos \& de la 
     Fuente Marcos (2014, 2016c). The conclusion of this study was that the known quasi-satellites form a very heterogeneous, dynamically 
     speaking, transient group. This result suggests that they have been temporarily captured from the Earth-crossing small-body 
     populations. If we compare the results obtained here with those in de la Fuente Marcos \& de la Fuente Marcos (2014, 2016c), that 
     conclusion is only confirmed although the orbital evolution of (164207)~2004~GU$_{9}$ clearly resembles that of (469219) 2016~HO$_{3}$, 
     even if their orbital inclinations are somewhat different. Asteroid 469219 is also significantly more stable than 164207 as it remains 
     for a longer period of time inside Earth's co-orbital zone. The quality of the orbital solutions of both objects is comparable, 
     therefore their different level of dynamical stability is a robust, distinctive feature. In principle, it may appear somewhat 
     surprising being able to find so much orbital diversity among transient Earth quasi-satellites however, as pointed out above (see 
     Section 3), the overlapping of a multiplicity of secular resonances makes this region particularly chaotic. 

     The effect of secular resonances on the dynamics of minor bodies following orbits with $a$ smaller than 2~au and relatively low values 
     of $e$ was first studied by Michel \& Froeschl\'e (1997). These early results were further extended by Michel (1997, 1998). Michel \& 
     Froeschl\'e (1997) concluded that minor bodies with $0.9<a<1.1$~au are subjected to the Kozai-Lidov resonance that, at low inclination, 
     induces orcillation of the argument of perihelion around the values 0\degr or 180\degr, the second variant of the Kozai-Lidov 
     mechanism (see e.g. Michel \& Thomas 1996). This behaviour has been found for 2009 DA$_{43}$ (see Section 3.7), but 469219 and several 
     other co-orbitals are not currently affected by the Kozai-Lidov mechanism.  

     Asteroid 469219 exhibits a number of dynamical features that make it unusual. In addition, it is not often that an Earth co-orbital as 
     small as this one ---that encounters our planet at relatively large distances--- sports an orbital determination as precise as the one 
     in Table \ref{elements}. This is possible because, in terms of average geocentric distance, it is one of the closest known co-orbitals 
     if not the closest. Earth quasi-satellites have not been the target subject of specific surveys. For instance, the discovery of 
     2014~OL$_{339}$ (Vaduvescu et al. 2014, 2015) was the by-product of a standard NEO survey, EURONEAR (Vaduvescu et al. 2008); the others 
     also were serendipitous, not planned, findings. Therefore, the identification by chance in less than a year of two objects moving in 
     similar orbits, 2015~SO$_{2}$ and 469219, that also exhibit matching orbital evolution within several thousand years backwards and 
     forward in time, strongly suggests that these two and other ---yet to be discovered--- NEOs may be part of a group of dynamical origin 
     as the ones described in e.g. de la Fuente Marcos \& de la Fuente Marcos (2016d).

     Asteroid 469219 cannot be considered part of the group of NEOs moving in Earth-like orbits or Arjunas (de la Fuente Marcos \& de la 
     Fuente Marcos 2013, 2015a,c) because its eccentricity is perhaps too high. However, it shares a number of dynamical features with them,
     in particular those derived from switching between co-orbital states.

     Our calculations show that 469219 is unlikely to collide with our planet. Its orbital properties are such that it always remains
     at a safe distance from the Earth but still close enough to make it an attractive target for future in situ study. Its relative 
     velocity at encounter remains in the range $\sim$3 to 5 km s$^{-1}$ for approaches in the near future.

  \section{Conclusions}
     In this paper, we have explored the orbital evolution of the recently discovered NEO (469219) 2016~HO$_{3}$. This study has been 
     performed using $N$-body simulations. In addition, a number of dynamical issues regarding the stability of Earth's co-orbital asteroid 
     population have been re-examined. Our conclusions can be summarized as follows.
     \begin{enumerate}[(i)]
        \item Asteroid 469219 is an Earth co-orbital, the fifth known quasi-satellite of our planet and the smallest. Its present 
              quasi-satellite dynamical state started nearly 100 yr ago and it will end in about 300 yr from now, transitioning to a 
              horseshoe state. It is the closest known Earth quasi-satellite, in terms of average distance from our planet.
        \item Extensive $N$-body simulations show that it can be counted among the most stable known Earth co-orbitals, with a Lyapunov time
              close to 7500 yr. This object may stay within Earth's co-orbital zone for a time interval well in excess of 1 Myr.
        \item Like a few other known Earth co-orbitals, 469219 experiences repeated transitions between the quasi-satellite and 
              horseshoe dynamical states. Jupiter plays a major role in the operation of the dynamical mechanism responsible for these 
              transitions. 
        \item Although a few other NEOs move in orbits similar to that of 469219, only one of them ---2015 SO$_{2}$--- exhibits a
              dynamical evolution that closely resembles that of the object discussed here.
        \item Our orbital analysis singles out 469219 as a very suitable candidate for spectroscopic studies as it will remain well 
              positioned with respect to our planet for many decades. Its location within the NEO orbital parameter space makes it also an
              attractive target for future in situ study.
        \item Exploratory calculations show that the assessment of the stability of Earth co-orbitals following paths similar to those of 
              2015 SO$_{2}$ and 469219 depends on the uncertainty in the value of the mass of Jupiter. In general, an improved 
              determination of the value of the Jovian mass will translate into a more robust statistical evaluation of the stability of 
              Earth co-orbitals.  
     \end{enumerate}

  \section*{Acknowledgements}
     We thank the anonymous referee for his/her constructive and particularly helpful report, S.~J. Aarseth for providing the code used in 
     this research and for his helpful comments on the effect of the uncertainties in the masses of the giant planets on the results of 
     $N$-body simulations and on earlier versions of this work, R. A. Jacobson for providing the JUP310 ephemeris and for valuable comments
     on the uncertainty in the value of the Jovian mass, and S. Deen for finding precovery images of (469219) 2016~HO$_{3}$ that improved 
     the orbital solution of this object significantly and for additional comments. In preparation of this paper, we made use of the NASA 
     Astrophysics Data System, the ASTRO-PH e-print server, the MPC data server and the NEODyS information service.

  \bsp
  \label{lastpage}
\end{document}